\documentclass[11pt]{article}
\usepackage{ amsmath, amsthm, amssymb } 
\usepackage{adjustbox}
\usepackage{color} 
\usepackage{subfig}
\usepackage{cite}
\usepackage{graphicx}
\newcommand{\be} { \begin{equation} } 
\newcommand{\ee} { \end{equation} } 
\newcommand{\nn}{\nonumber \\}

\newcommand{\M}{M}
\newcommand{\D}{\mathfrak{D}}
\newcommand{\I}{\mathcal{I}}
\newcommand{\J}{\mathcal{J}}
\newcommand{\Hh}{\mathcal{H}}
\newcommand{\K}{\mathcal{K}}
\newcommand{\N}{\mathbb{N}}

\newcommand{\EK}{{\rm K}}
\newcommand{\EE}{{\rm E}}

\newcommand{\Bub}{{\rm Bub}}

\newcommand{\Cut}{{\rm Cut}}
\newcommand{\e}{\epsilon}
\newcommand{\Tr}{{\rm  Tr}}
\newcommand{\C}{\mathcal{C}}
\newcommand{\eps}{\epsilon}

\allowdisplaybreaks

\begin{document} 
\setlength{\unitlength}{1.3cm} 
\begin{titlepage}
\vspace*{-1cm}
\begin{flushright}
TTP17-021
\end{flushright}                                
\vskip 3.5cm
\begin{center}
\boldmath
 
{\Large\bf Maximal cuts and differential equations for Feynman integrals. \\
An application to the three-loop massive banana graph\\[3mm] }
\unboldmath
\vskip 1.cm
{\large Amedeo Primo}$^{a,b,}$
\footnote{{\tt e-mail: amedeo.primo@pd.infn.it}} and
{\large Lorenzo Tancredi}$^{c,}$
\footnote{{\tt e-mail: lorenzo.tancredi@kit.edu}} 
\vskip .7cm
{\it $^a$ Dipartimento di Fisica ed Astronomia, Universit\`a di Padova, Via Marzolo 8, 35131 Padova, Italy } \\
{\it $^b$ INFN, Sezione di Padova, Via Marzolo 8, 35131 Padova, Italy} \\
{\it $^c$ Institute for Theoretical Particle Physics, KIT, 76128 Karlsruhe, 
Germany } 
\end{center}
\vskip 2.6cm

\begin{abstract}
We consider the calculation of the master integrals of the 
three-loop massive banana graph. In the case of equal internal masses, the graph
is reduced to three master integrals which satisfy an irreducible system of three coupled linear
differential equations.  The solution of the system
requires finding a $3 \times 3$ matrix of homogeneous solutions. We show how the maximal
cut can be used to determine all entries of this matrix in terms of products of elliptic
integrals of first and second kind of suitable arguments. All independent solutions are
found by performing
the integration which defines the maximal cut on different contours.
Once the homogeneous solution is known, the inhomogeneous solution
can be obtained by use of Euler's variation of constants.

\end{abstract}
\vfill
\end{titlepage}                                                                
\newpage

\section{Introduction}
\label{sec:intro} \setcounter{equation}{0} 
\numberwithin{equation}{section} 

The study of the mathematical structures that characterize multiloop Feynman integrals
has played a crucial role for the most recent developments in the computation of higher
order radiative corrections for many processes of direct phenomenological interest
at the LHC.
The discovery of integration by parts identities~\cite{Tkachov:1981wb,Chetyrkin:1981qh}
together with the differential equations method~\cite{Kotikov:1990kg,Bern:1993kr,Remiddi:1997ny},
in particular as extensively applied to compute entire families of massless four-point Feynman 
integrals~\cite{Gehrmann:1999as}, allowed to successfully deal with large classes of problems
that had remained out of reach before.
One step further has been made 
with the introduction of the concept of a canonical basis~\cite{Henn:2013pwa}.
The latter allows to render the computation of multiloop Feynman integrals, 
which can be expressed in terms of multiple 
polylogarithms~\cite{Goncharov,Remiddi:1999ew,Vollinga:2004sn}, 
almost entirely algorithmic. In the last years, different methods for the construction of a canonical basis have been proposed, according to
the specific properties of the differential equations, such as a linear dependence on the spacetime dimensions \cite{Argeri:2014qva} or the dependence on a single kinematic 
variable~\cite{Lee:2014ioa,Ablinger:2015tua,Lee:2016lvq} and concrete steps towards algorithms also valid for multi-scale problems 
have been made in~\cite{Gehrmann:2014bfa,Meyer:2016slj,Adams:2017tga}.
The striking simplicity highlighted by the use of canonical bases can be traced back to the rich algebraic structures which can be associated to multiple polylogarithms~\cite{Goncharov:2001iea,Goncharov:2010jf,Duhr:2011zq,Duhr:2012fh}.
More recently, a big effort has been devoted in studying how to possibly extend some
of these structures to include more complicated Feynman integrals, which cannot be evaluated
in terms of multiple polylogarithms only. These new mathematical functions
start to appear at the two-loop order, in particular in the presence of internal massive 
propagators~\cite{Broadhurst:1987ei,Bauberger:1994by,Bauberger:1994hx,Laporta:2004rb,Aglietti:2007as,CaronHuot:2012ab,Bloch:2013tra,Adams:2013nia,Remiddi:2013joa,Adams:2014vja,Adams:2015gva,Adams:2015ydq,Remiddi:2016gno,Adams:2016xah,Bonciani:2016qxi,vonManteuffel:2017hms}.

A way to understand the jump in complexity entailed by these integrals is
to look at the details of their computation with the differential equations method.  
From this point of view, Feynman integrals which evaluate to multiple polylogarithms are found to
satisfy first order linear differential equations in the external invariants, at least 
in the limit of even space-time dimensions, $d=2\,n$, $n \in \N$\footnote{We recall here that the Laurent coefficients 
of a Feynman integral expanded around $d=4$ space-time dimensions can be recovered
from the corresponding expansion around any even number of dimensions~\cite{Tarasov:1996br}.
A way to determine whether the differential equations can be decoupled in $d=2 n$ space-time
dimensions is described in~\cite{Tancredi:2015pta}.}. 
What this means is that,
it is always possible to write an integral representation (by simple quadrature of the equations)
for the Laurent coefficients of their series expansion in $\epsilon = (4-d)/2$. 
The situation changes completely when we consider graphs that cannot be expressed in terms of
multiple polylogarithms. In this case, the corresponding master integrals are found to satisfy irreducible
higher-order differential equations even in the limit $d=4$, whose most famous example is 
undoubtedly the second-order differential
equation satisfied by the two-loop massive sunrise graph~\cite{Laporta:2004rb}.

As it is well known, an integral representation for the solution of a higher-order differential equation
can be found only if the full set of its
homogeneous solutions is known. 
Unfortunately, given a generic higher-order differential equation
with rational coefficients, except for a limited number of special cases, there is no way to
determine the full set of its homogeneous solutions. This constituted for a long time a
serious problem in the computation of any Feynman integral of this type.
More recently a new picture started to emerge. We showed in~\cite{Primo:2016ebd} that
an integral representation for (one of) the homogeneous solutions of the differential equation
satisfied by any Feynman integral can be obtained by computing its maximal cut. 
A similar idea was proposed in~\cite{Lee:2012te} in the context of the DRA method.
The statement is independent on the number of space-time dimensions $d$ we are
interested in, in the sense that the maximal cut computed in
any, continuous, number of dimensions $d$ provides an homogeneous solution for its
$d$-dimensional differential equation. Specializing to the physically interesting
case of $d=4$, one obtains at once the homogeneous solution of the corresponding 
$4$-dimensional differential equation. 
It was then shown in~\cite{Frellesvig:2017aai}, that the computation
of the maximal cut can be simplified by the use of the so-called
Baikov representation~\cite{Baikov:1996rk}.
Moreover, the importance of the unitarity cuts for understanding more general mathematical structures
of Feynman integrals has been pointed out recently in~\cite{Abreu:2015zaa,Abreu:2017ptx,Abreu:2017enx},
while in~\cite{Zeng:2017ipr} it was shown
how to use cuts to derive efficiently the differential equations.

In spite of the increasing effort, until today only a limited number of examples have been 
considered in the literature. Interestingly, all examples considered could always be reduced to
differential equations of at most degree two,
whose homogeneous part could always
be solved in terms of complete elliptic integrals of first and second kind.
Quite in general,
for any second-order differential equation, once one homogeneous solution is known,
the second can be obtained by a simple quadrature.
The same is not true for higher-order equations.
Therefore, a very important issue, that was not fully investigated in~\cite{Primo:2016ebd}, is
whether it is possible to obtain \textsl{all} independent homogeneous solutions from the computation
of the maximal cut only. This becomes crucial if one is interested in 
Feynman integrals which fulfil third- or higher-order differential equations.

In this paper we address this problem and show that the answer to this question is indeed affirmative
and that all independent homogeneous solutions can be obtained by evaluating the 
maximal cut along different independent contours, which do not cross any branch cut of the
integrand. We notice here that there is a clear correspondence between this simple idea 
and the methods described in~\cite{Lee:2013hzt} to count the number of independent master integrals.\footnote{Indeed, if a Feynman integral is reduced to $N$ master integrals in $d$ dimensions,
the latter will satisfy a system of $N$ coupled differential equations. A complete solution of the latter
would therefore require to find $N$ independent homogeneous solutions.}
While this can be seen very easily already in the case of the two-loop massive sunrise graph,
here we move one step forward and consider the first example of a Feynman graph
that fulfils an irreducible third-order differential equation: the three-loop massive banana graph. 
This calculation requires finding three independent homogeneous solutions, for which we show how to
derive integral representations from the study of its maximal cut only. 
We show, moreover, that these integrals can be evaluated explicitly in terms of products of elliptic integrals
of first and second kind of complicated arguments. This result, which is very non-trivial,
was expected. 
In fact, the third-order differential equation satisfied by this graph has been studied long ago 
by G.S. Joyce in the context of cubic lattice Green functions~\cite{Joyce1}, 
where it was shown that this equation is a so-called \textsl{symmetric square} and its solution
can be therefore expressed as products of solutions of a lower second-order differential equation.
Finally, we recall the reader that an alternative calculation of this graph has been presented 
in~\cite{Bloch:2014qca},
where it was shown that the three-loop banana graph in $d=2$ can be written as an
\textsl{elliptic three-logarithm}.

The rest of the paper is organized as follows. In Section~\ref{sec:sunrise} we reconsider in detail
the well known
case of the two-loop massive sunrise graph, showing how the two independent homogeneous
solutions of its
second-order differential equation
can be found by integrating the maximal cut over the only two independent contours that one can build
without crossing any branch cut of the integrand.
We switch then to the more interesting case of the three-loop massive banana graph
in Section~\ref{sec:banana}, where we derive the system of three-coupled differential equations
that it satisfies.
In Section~\ref{sec:homsys} we focus on the system of differential equations around $d=2$ 
space-time dimensions, which is known to be equivalent to the corresponding
expansion close to $d=4$\cite{Tarasov:1996br}. 
We compute the maximal cut of the three-loop banana graph in
Section~\ref{sec:maxcuthomsol} and show that by evaluating it along different contours we obtain
at once all independent solutions of the homogeneous differential equations.
In Section~\ref{sec:symmsquare} we show that the same third-order differential equation
is a symmetric square, we describe its solution as derived in~\cite{Joyce1} and
discuss its equivalence to ours. Finally in Section~\ref{sec:inhomsol} we put everything together
and compute the inhomogeneous solution, continuing it analytically to the whole phase space.
We also provide some appendices where we describe some of the technical calculations
required  in the main text.\newline

\noindent {\Large \textbf{Note Added}}

\noindent
Right before the completion of this work an interesting paper appeared on the arXiv~\cite{Bosma:2017ens}, 
containing similar findings to the one presented here.
In~\cite{Bosma:2017ens} it is shown, with many explicit examples,
how to build the full set of independent homogeneous
solutions of the differential equations by integrating the maximal cut on different regions,
for generic values of the space-time dimensions $d$. This is done efficiently using Baikov representation.
While our conclusions are similar, in our paper we do not dwell on general methods for an efficient computation of the cut 
in any numbers of dimensions, as we are primarily interested in the calculation of the Laurent 
coefficients of the master integrals close to $d = 4$ and the complexity of this calculation 
is determined entirely by the value of the homogeneous solutions at $d=2 n$, $n \in \N$.
Instead,
we apply  for the first time these ideas to solve a system of three coupled differential equations,
which require a generalization of the methods used for the two-loop massive sunrise graph
and other similar examples.
From this point of view, we believe our results are complementary to the ones presented 
in~\cite{Bosma:2017ens}.

\section{Revisiting the two-loop massive sunrise graph}
\label{sec:sunrise} \setcounter{equation}{0} 
\numberwithin{equation}{section} 

Before looking in detail at the case of the three-loop massive banana graph, it is useful to
reconsider the better understood case of the two-loop massive sunrise.
This will help us to illustrate some of the techniques used later for the three-loop banana graph
in a simpler environment.

As it is well known, the sunrise graph satisfies a coupled system of
 two linear differential equations~\cite{Laporta:2004rb};
its homogeneous solution is therefore a $2 \times 2$ matrix, whose entries are linear combinations of
square-roots and complete elliptic integrals of first and second kind, see for example~\cite{Remiddi:2016gno}. 
The entries of this matrix were determined for the first time in~\cite{Laporta:2004rb} by studying
the imaginary part of the graph.
Taking inspiration from this calculation, in~\cite{Primo:2016ebd} we showed that the computation of
the maximal cut of any family of Feynman integrals naturally generalizes the approach 
of~\cite{Laporta:2004rb}
and allows one to determine an integral representation for an homogeneous solution of its system
of differential equations irrespective of its degree.
As argued in~\cite{Primo:2016ebd}, if an homogeneous solution can be determined
in terms of complete elliptic integrals, it is then straightforward to find the second solution using the
properties of the elliptic integrals. While this is very useful, it cannot be straightforwardly extended
to more general cases which involve higher-order differential equations and are not expected to
be solved in terms of elliptic integrals only. 

The goal of this section is to show how the study of the maximal cut of the sunrise graph provides 
at once \textsl{both independent solutions}. 
While the validity of this claim is somehow obvious, its consequences are non-trivial and far-reaching.
This remains valid, in fact, for any Feynman integral and provides therefore an effective method
for constructing integral representations for \textsl{all independent homogeneous solutions}
of the differential equations satisfied by the graph, regardless of their complexity.
In particular, we will show explicitly how this is done by integrating the maximal cut of the sunrise graph 
along the only two linearly independent integration contours which do not cross any singularity of the integrand.

Let us consider the equal-mass sunrise graph 
\begin{align}
&  \adjustbox{valign=m}{\includegraphics[scale=0.8,trim=0 0.2cm 0 0cm]{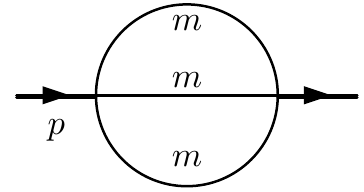} }
 \hspace{-0.2cm} = S(d;u)  =  \int \, 
\frac{ \D^dk_1\, \D^d k_2}{[k_1^2-m^2] [k_2^2-m^2]
[(k_1-k_2-p)^2-m^2] }\,,   \label{eq:sun}
\end{align}
 where $u= p^2/m^2$. The graph in $d=2$ space-time dimensions 
 satisfies the following second-order differential equation
\be
\left[ \frac{d^2}{du^2} + \left( \frac{1}{u} + \frac{1}{u-1} + \frac{1}{u-9} \right) \frac{d}{du}
+ \left( -\frac{1}{3u} + \frac{1}{4(u-1)} + \frac{1}{12(u-9)}  \right) \right] S(u) = 0\,, \label{eq:deqsun}
\ee
where we neglected the inhomogeneous terms which are irrelevant here and we set $S(2;u) =S(u)$.
As it is well known, the maximal cut of the sunrise graph  in $d = 2$ can be written as
\be
\Cut\left( S(u)\right) = \oint_\C \frac{db}{\sqrt{\pm b\left( b-4 \right) \left( b- (\sqrt{u}-1)^2 \right) 
\left( b- (\sqrt{u}+1)^2 \right)}} = \oint_\C \frac{db}{\sqrt{\pm R_4(b,u)}}\,, \label{eq:cutsun}
\ee
where we use the notation $\Cut(S(u))$ 
for the maximal cut of $S(u)$ and 
we have not fully specified neither the integration contour $\C$ nor the sign of the argument
of the root.
We claim that the integration along any contour $\C$ which does not cross any branching point
of the integrand produces a solution of~\eqref{eq:deqsun}. 
In particular, we will see that there are only two possible independent contours of such type and that by integrating along them we get at once
both independent solutions of~\eqref{eq:deqsun}.

\begin{figure}
  \centering
  \subfloat[][]{%
    \includegraphics[width=0.50\textwidth,trim=0 0 0 0.27cm]{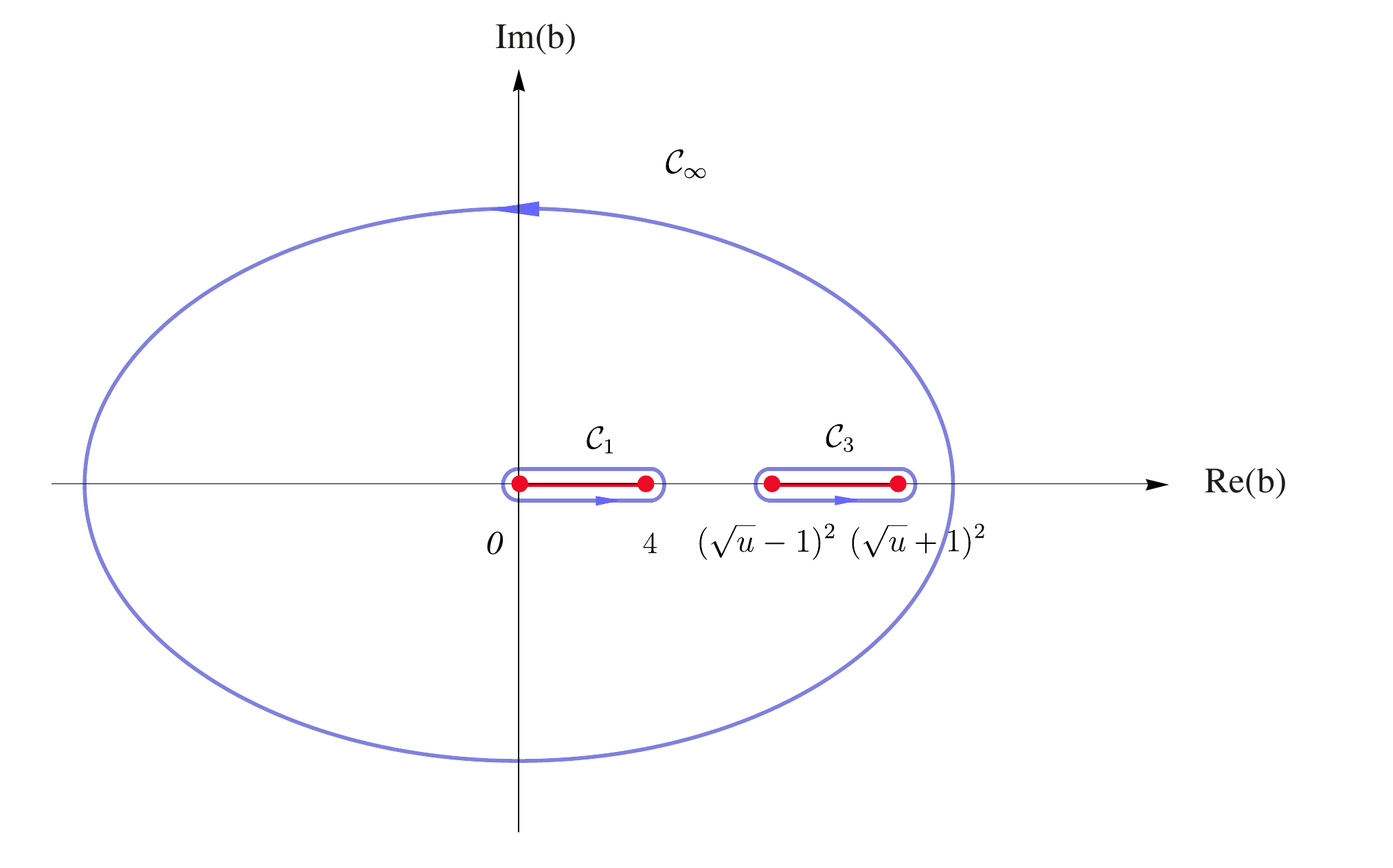}
    \label{fig:contoursun1} }
  \subfloat[][]{%
    \includegraphics[width=0.50\textwidth,trim=0 0 0 0.27cm]{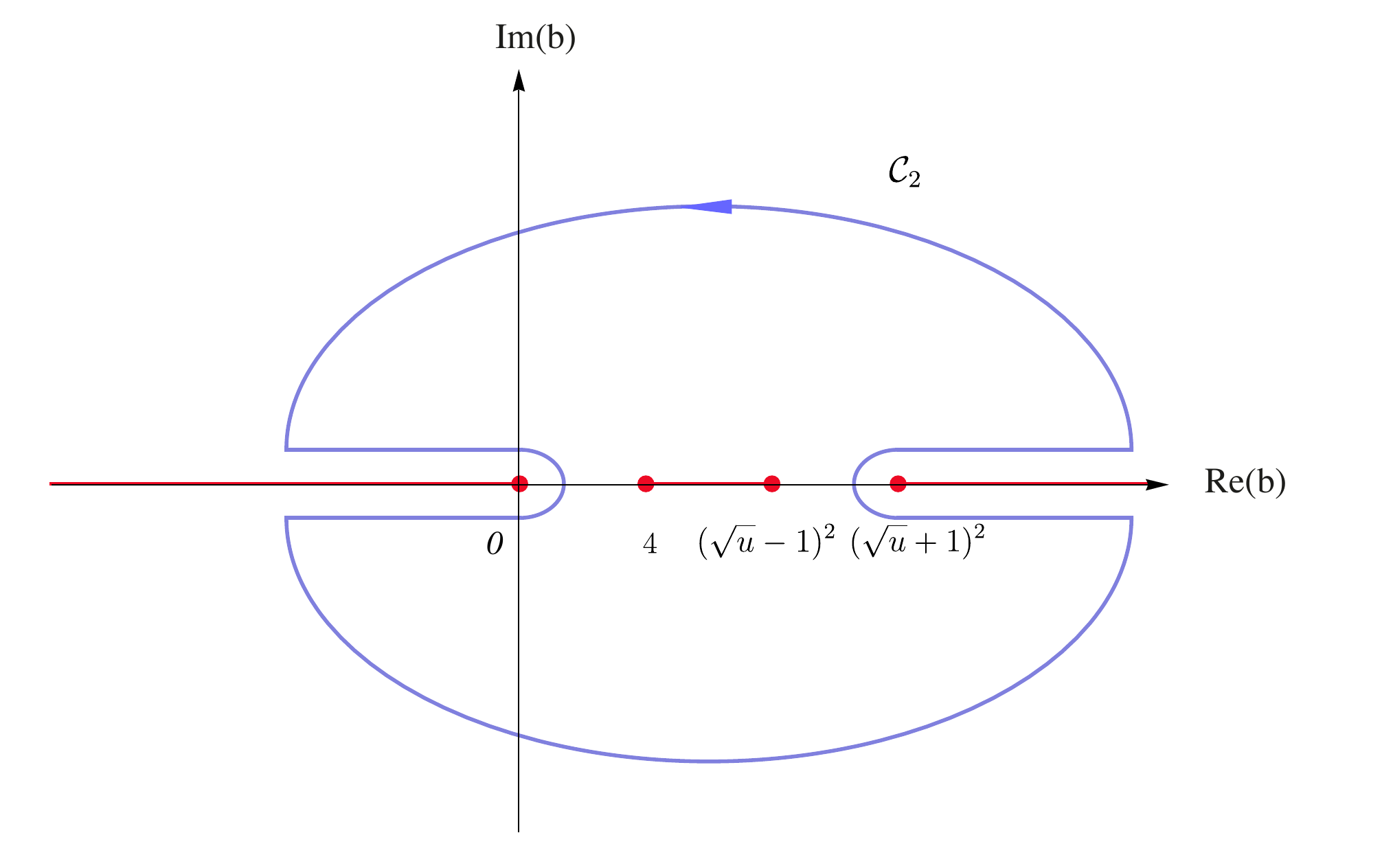}
      \label{fig:contoursun2}
  }
  \caption{Left panel: The contours $\C_1$, $\C_3$ and $\C_\infty$. The branches of the integrand for the \textsl{positive sign} in the root in Eq.~\eqref{eq:cutsun} are drawn
    in red. Right panel: The contour $\C_2$. The branches of the integrand for the \textsl{negative sign} in the root in Eq.~\eqref{eq:cutsun} are drawn in red.}
  \label{fig:contours}
\end{figure}

First of all, the square-root has four branching points. By choosing $u > 9$
 we have 
 \be
 0 < 4 < (\sqrt{u}-1)^2 < (\sqrt{u}+1)^2 \,. \label{eq:rootsun}
 \ee
 The ordering of the branching points depends on the value of $u$, 
 but the argument used below does not depend on it.
Given the four branching points it should be obvious that, depending on the sign that 
we pick in~\eqref{eq:cutsun}, there are four possible integration contours
which we can draw without crossing the branch cuts.
If we choose the plus sign, the integrand develops a branch cut for
$0<b<4$ and  $(\sqrt{u}-1)^2<b<(\sqrt{u}+1)^2$.
If we pick the minus sign the branches are for $-\infty < b <0$,
$4<b<(\sqrt{u}-1)^2$ and $(\sqrt{u}+1)^2 < b < +\infty$.
In the first case, i.e. picking a plus sign, we can clearly draw the two contours $\C_1$ and $\C_2$ 
depicted in 
Figure~\ref{fig:contoursun1}. The third contour, $\C_\infty$, is instead equivalent to the sum of the two, 
and we will need it later on. 
In the second case, we can draw instead only one single contour,
see Figure~\ref{fig:contoursun2}, giving a total of three apparently different possibilities.

Nevertheless, it is easy to see~\cite{Laporta:2004rb}
that two of the three contours are equivalent.
Let us consider the following integral
\be
\oint_{\C_\infty} \frac{db}{\sqrt{R_4(b,u)}} = 0\,, \label{eq:cinf1}
\ee
where $\C_\infty$ is the contour at infinity encircling the four roots~\eqref{eq:rootsun}
depicted in Figure~\ref{fig:contoursun1}.
Clearly the integral is zero since the contour does not contain any poles and
the integrand goes as $1/b^2$ when $b \to \infty$.
On the other hand, with the sign of the root in~\eqref{eq:cinf1}, the integrand
has two cuts, one for $0<b<4$ and the other for $(\sqrt{u}-1)^2<b<(\sqrt{u}+1)^2$.
We can then imagine to shrink $\C_\infty$ to encircle the two branch cuts and we get
from~\eqref{eq:cinf1}
\be
\oint_{\C_\infty} \frac{db}{\sqrt{R_4(b,u)}} = \oint_{\C_1}\frac{db}{\sqrt{R_4(b,u)}} 
+ \oint_{\C_3}\frac{db}{\sqrt{R_4(b,u)}} = 0\,, 
\ee
which proves that the two integrals are not independent.
In this way we are left with two independent contour integrals which provide precisely
the two independent solutions of~\eqref{eq:deqsun}, say $\C_1$ and $\C_2$.
Now, by shrinking $\C_1$, $\C_2$ and $\C_3$ on the corresponding branch cuts one finds
an equivalent representation as one-dimensional real integrals
\begin{align}
\oint_{\C_1} \frac{db}{\sqrt{R_4(b,u)}} = 2\, i\, &\int_{0}^4 \frac{db}{\sqrt{-R_4(b,u)}}\,, \qquad
\oint_{\C_2} \frac{db}{\sqrt{-R_4(b,u)}} = 2\,i\,\int_{4}^{(\sqrt{u}-1)^2} \frac{db}{\sqrt{R_4(b,u)}} \nonumber \\
&\oint_{\C_3} \frac{db}{\sqrt{R_4(b,u)}} =  - 2\, i\, \int_{(\sqrt{u}-1)^2}^{(\sqrt{u}+1)^2} \frac{db}{\sqrt{-R_4(b,u)}} \,,
\label{eq:contours}
\end{align}
where the sign of the roots on the right hand side of the formulas is chosen to deal with real 
integrals. Alternatively, the contour $\C_2$ can also be sent to infinity providing a second
integral representation for the integral
\begin{align}
\oint_{\C_2} \frac{db}{\sqrt{-R_4(b,u)}} = 2\,i \,
\left( \int_{-\infty}^0 + \int_{(\sqrt{u}+1)^2}^{+\infty} \right) \frac{db}{\sqrt{R_4(b,u)}}\,.
\label{eq:contour2}
\end{align}
We have therefore determined two independent solutions as the integrals over the two independent
contours, say $\C_1$ and $\C_2$, which in turn we can written as one-dimensional real integrals 
as follows
\begin{align}
&J(u) = \int_{0}^4 \frac{db}{\sqrt{-R_4(b,u)}} = \int_{(\sqrt{u}-1)^2}^{(\sqrt{u}+1)^2} \frac{db}{\sqrt{-R_4(b,u)}}\,,
\nonumber \\ 
&I(u) = \int_{4}^{(\sqrt{u}-1)^2} \frac{db}{\sqrt{R_4(b,u)}} = \left( \int_{-\infty}^0 + \int_{(\sqrt{u}+1)^2}^{+\infty} \right) \frac{db}{\sqrt{R_4(b,u)}}\,.
\end{align}
It was shown in~\cite{Laporta:2004rb,Remiddi:2016gno} that these functions are indeed
the two independent solutions of~\eqref{eq:deqsun}.
\newline

Clearly, the analysis we performed is completely general and does not depend on
the precise position of the branching points~\eqref{eq:rootsun}. In fact, given a generic 
integral of the form
\be
I_{\C_a} = \oint_{\C_a} 
\frac{da}{\sqrt{(a-a_1) \,(a-a_2)\, 
\left(a - a_3 \right) \left( a -a_4\right)}} = 
 \oint_{\C_a} \frac{da}{\sqrt{R(a,a_1,a_2,a_3,a_4)}} \,,
\ee
where $a_1$, ..., $a_4$ are four distinct roots such that
$$a_1<a_2<a_3<a_4$$
the same considerations apply and one
finds that there are only two independent contours which are equivalent to
the following 
one dimensional real integrals
\begin{align}
&I(a_1,a_2) = \int_{a_1}^{a_2} \frac{da}{\sqrt{-R(a,a_1,a_2,a_3,a_4)}} = 
 \int_{a_3}^{a_4} \frac{da}{\sqrt{-R(a,a_1,a_2,a_3,a_4)}}\,,  \nonumber \\
&I(a_2,a_3) = \int_{a_2}^{a_3} \frac{da}{\sqrt{R(a,a_1,a_2,a_3,a_4)}} = 
\left(  \int_{-\infty}^{a_1} + \int_{a_4}^{+\infty}  \right)\frac{da}{\sqrt{R(a,a_1,a_2,a_3,a_4)}} \,.
\label{eq:contell}
\end{align}
As for the explicit case of the sunrise with equal masses, we chose the sign in the root
in order to deal with real integrals everywhere.

As a last remark, the integrals in Eqs.~\eqref{eq:contell} are nothing but complete
elliptic integrals of the first kind.
To see this, we perform the two standard changes of variables for the two integrals respectively
\begin{align}
&I(a_1, a_2) \quad \longrightarrow \quad t^2 = \frac{(a_4-a_2)(a-a_1)}{(a_2-a_1)(a_4-a)} \,,\nonumber 
 \\
&I(a_2, a_3) \quad \longrightarrow \quad t^2 = \frac{(a_1-a_3)(a-a_2)}{(a_3-a_2)(a_1-a)}\,, \label{eq:chvar}
\end{align}
and obtain
\begin{align}
& I(a_1,a_2) = \frac{2}{\sqrt{(a_3-a_1)(a_4-a_2)}}
\EK\left( w_1\right)\,, \\
& I(a_2,a_3) = \frac{2}{\sqrt{(a_3-a_1)(a_4-a_2)}}
\EK\left( 1-w_1\right)
\,,
\end{align}
where 
\be
\EK(w) = \int_0^1 \frac{dz}{\sqrt{(1-z^2)(1-w\,z^2)}}\,\quad \mbox{with} \quad \Re{(w)} < 1\,,
\ee
is the elliptic integral of the first kind and
\begin{align}
w_1 = \frac{(a_2-a_1)(a_4-a_3)}{(a_3-a_1)(a_4-a_2)} \,.
\end{align}
Indeed, a standard result of the theory of the complete elliptic integrals shows that 
$K(w)$ and $K(1-w)$ satisfy the same second-order differential equation, of which
they constitute the two independent solutions.

The analysis carried out in this section might seem somewhat redundant, as the
theory of the elliptic integrals has been very well understood for a long time.
Nevertheless, when considering the three-loop banana graph, we will see that many of the ideas and of the results derived here can be directly borrowed or trivially extended to more complicated cases. 
We believe that  this will make our analysis in this much less trivial case, much more transparent.

\section{The three-loop massive banana graph}
\label{sec:banana} \setcounter{equation}{0} 
\numberwithin{equation}{section} 

We consider the three-loop two-point integral family defined by
\begin{align}
&  \adjustbox{valign=m}{\includegraphics[scale=1.,trim=0 0 0 -0.27cm]{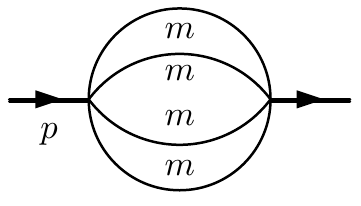} }
= I_{a_1,a_2,a_3,a_4,a_5,a_6,a_7,a_8,a_9} \Big|_{a_5,\cdots\,, a_9 < 0} \nonumber \\
& = \int \, 
\frac{ \D^dk_1\, \D^d k_2 \D^d k_3 \;\; (k_3^2)^{-a_5}(k_1\cdot p)^{-a_6}(k_2\cdot p)^{-a_7}(k_3\cdot p)^{-a_8}(k_1\cdot k_2)^{-a_9}}{[k_1^2-m^2]^{a_1} [k_2^2-m^2]^{a_2} [(k_1-k_3)^2-m^2]^{a_3}
[(k_2-k_3-p)^2-m^2]^{a_4} }\,,   \label{eq:topo}
\end{align}
with  $p^2  = s  \neq 0 $. Our integration measure is defined as
\be
\int \D^d k = \frac{(m^2)^{\frac{4-d}{2}}}{i \pi^{d/2}\Gamma\left(\frac{4-d}{2}\right)}
\int \frac{d^d k}{(2 \pi)^d}\label{eq:measure}
\ee
such that the one-loop tadpole integral reads
\be
\int \frac{ \D^d k }{k^2 - m^2} = \frac{2m^2}{d-2}\,. \label{eq:tadpole}
\ee

The family of integrals in~\eqref{eq:topo} with $a_5,...,a_9 < 0$
is often referred to as three-loop banana graph or three-loop sunrise graph.
A first step towards the computation of these integrals is to use integration by 
parts identities to reduce them to a basis of master integrals.
In this case, this step can be very easily achieved with any of the publicly available
reduction codes like, for example, Reduze 2~\cite{vonManteuffel:2012np}.
There is only one sub-topology, the three-loop massive tadpole, which is reduced to one
single master integral. The top-topology, instead, in the limit of equal internal masses  
can be reduced to three independent master integrals.
The choice of the master integrals is of course arbitrary. 
Since massive two-point functions are IR finite, they are best studied in the vicinity of $d=2$ instead
of $d=4$, since for $d = 2$ the scalar amplitude is also UV finite.
We define therefore $\epsilon = (2 - d)/2$ and we pick the following basis
\begin{align}
\I_{1}(\e;s) =&(1+2\e)(1+3\e)(m^2)^{-2} I_{1, 1, 1, 1, 0, 0, 0,0,0}\,,\nn
\I_{2}(\e;s) =&(1 + 2 \e)(m^2)^{-1} I_{2, 1, 1, 1, 0, 0, 0,0,0}\,,\nn
\I_{3}(\e;s) =&I_{2, 2, 1, 1, 0, 0, 0,0,0}\,,
\label{eq:basis2}
\end{align}
where the normalization factors have been chosen for later convenience.
For the three-loop tadpole we chose instead the master integral $I_{2, 2, 2, 0 ,0, 0, 0,0,0}$,
which in our normalization becomes simply
\be
\I_0(\e;m^2) = I_{2, 2, 2, 0 ,0, 0, 0,0,0} = 1\,. \label{eq:tadpole}
\ee

The dependence of the master integrals  on the momentum
transfer $p^2 = s$ can be conveniently parametrized in terms of the dimensionless variable
\begin{align}
x=\frac{4m^2}{s}\,.
\end{align}
The system of differential equations in $x$ satisfied by $\I_1$, $\I_2$ and $\I_3$ is given by
\begin{align} 
\frac{d}{dx} \left(\begin{matrix} \I_{1}(\e;x) \\ \I_{2}(\e;x)\\ \I_{3}(\e;x) \end{matrix} \right) =&
B(x) \left(\begin{matrix} \I_{1}(\e;x) \\ \I_{2}(\e;x) \\ \I_{3}(\e;x) \end{matrix} \right) 
+ \epsilon\, D(x)\left(\begin{matrix} \I_{1}(\e;x) \\ \I_{2}(\e;x)\\ \I_3(\e;x) \end{matrix} \right)
+\left(\begin{matrix} 0 \\ 0\\ -\frac{1}{2 (4 x-1)} \end{matrix} \right)
\, ,
 \label{eq:topsys2}
\end{align}
where $B(x)$ and $D(x)$ are $3 \times 3$ matrices that do not depend on $\epsilon$, 
\begin{align}
B(x) =&\left(
\begin{array}{ccc}
 \frac{1}{x} & \frac{4}{x} & 0 \\
 -\frac{1}{4 (x-1)} & \frac{1}{x}-\frac{2}{x-1} & \frac{3}{x}-\frac{3}{x-1} \\
 \frac{1}{8 (x-1)}-\frac{1}{8 (4 x-1)} & \frac{1}{x-1}-\frac{3}{2 (4 x-1)} &
   \frac{1}{x}-\frac{6}{4 x-1}+\frac{3}{2 (x-1)} \\
\end{array}
\right)\,,  \label{eq:matrixB} \\
D(x) = &\left(
\begin{array}{ccc}
 \frac{3}{x} & \frac{12}{x} & 0 \\
 -\frac{1}{x-1} & \frac{2}{x}-\frac{6}{x-1} & \frac{6}{x}-\frac{6}{x-1} \\
 \frac{1}{2 (x-1)}-\frac{1}{2 (4 x-1)} & \frac{3}{x-1}-\frac{9}{2 (4 x-1)} &
   \frac{1}{x}-\frac{12}{4 x-1}+\frac{3}{x-1} \\
\end{array}
\right)\,, \label{eq:matrixD}    \\                                
\end{align}
and the inhomogeneous term is proportional to the massive tadpole~\eqref{eq:tadpole}.
We stress once more that we study the solution of the differential equations in the vicinity of $d = 2$
and in our conventions $\epsilon = (2-d)/2\,.$

As it is easy to see, the structure of the system of differential equations is strikingly simple, 
as it is characterized by only four regular singular points $x= 0$, $x= 1/4$, $x=1$ and $x=\pm \infty$, which correspond, respectively, to
$s = \pm \infty $, $s = 16m^2$, $s=4 m^2$ and $s = 0$.
The structure of the singularities resembles closely that of the simpler two-loop massive
sunrise graph, see for example~\cite{Laporta:2004rb}.
Given the simplicity of the equations, it is particularly interesting to investigate the class of functions
that are needed for their solution and in which sense these functions generalize the
ones required for the integration of the two-loop sunrise graph. 
In the two-loop case, the homogeneous system of differential equations admits as
solutions complete elliptic integrals of the first and second kind.
Indeed, as we will see in the next section, the three-loop case
can be solved in terms of 
\textsl{products of complete elliptic integrals of first and second kind}.

\section{The homogeneous system}
\label{sec:homsys} \setcounter{equation}{0} 
\numberwithin{equation}{section} 
Before embarking in the solution of the differential equations, let us first recall which ingredients
are needed to solve a $3 \times 3$ coupled system, as the one in Eq.~\eqref{eq:topsys2}.
Since we are interested in calculating the solution as a Laurent expansion in $\e$,
the first step consists in solving the homogeneous system for $\e = 0$

\begin{equation}
\frac{d}{dx} \left(\begin{matrix} \I_{1H}(x) \\ \I_{2H}(x)\\ \I_{3H}(x) \end{matrix} \right) =
B(x) \left(\begin{matrix} \I_{1H}(x) \\ \I_{2H}(x) \\ \I_{3H}(x) \end{matrix} \right) \,,
\label{eq:syshom}
\end{equation}
where the suffix $H$ indicates that we are limiting ourselves to its homogeneous piece.
The solution of Eq.~\eqref{eq:syshom} requires finding a $3 \times 3$ matrix, say $G(x)$, defined 
such that

\begin{equation}
G(x) = \left(\begin{array}{ccc}
H_{1}(x) & J_{1}(x) & I_{1}(x) \\
H_{2}(x) & J_{2}(x) & I_{2}(x) \\
H_{3}(x) & J_{3}(x) & I_{3}(x) \\
\end{array}\right) \qquad \to \qquad 
\frac{d}{dx} G(x) = B(x)\, G(x)\,. \label{eq:Gmat}
\end{equation} 
The inverse of the matrix $G(x)$ is
\begin{align}
G^{-1}(x)=\frac{1}{W(x)}\left(
\begin{array}{ccc}
 I_{3} J_{2}-I_{2} J_{3} & I_{1} J_{3}-I_{3} J_{1} & I_{2} J_{1}-I_{1} J_{2} \\
 H_{3} I_{2}-H_{2} I_{3} & H_{1} I_{3}-H_{3} I_{1} & H_{2} I_{1}-H_{1} I_{2} \\
 H_{2} J_{3}-H_{3} J_{2} & H_{3} J_{1}-H_{1} J_{3} & H_{1} J_{2}-H_{2} J_{1} \\
\end{array}
\label{eq:Ginv}
\right)\,,
\end{align}
where $W(x)= \det\left(G(x)\right)$ is the Wronksian of the system~\eqref{eq:syshom}. 
As it is well known $W(x)$ satisfies the first order differential equation 
\begin{equation}
\frac{d}{dx} W(x) = \Tr \left ( B(x) \right) W(x)\,, \label{eq:Abel}
\end{equation}
which is often referred to as Abel's identity.
From the matrix $B(x)$ defined in Eq.~\eqref{eq:matrixB} we get at once
\begin{align}
\frac{d}{dx} W(x) = \frac{8 x^2-17 x+6}{2x (x-1)(4 x-1)} W(x)\,,
\end{align}
which admits the very simple solution
\begin{equation}
W(x) = \frac{c_1 x^3}{\sqrt{(1-4 x)^{3}(1-x)}}\,, \label{eq:Wronsk1}
\end{equation}
where we assumed for simplicity to work in the Euclidean region, $x < 0$.
The value of the integration constant $c_1$ can be determined only once the exact form 
of the solutions is known.

The simplicity of the Wronskian~\eqref{eq:Wronsk1} is a general feature 
and can be easily understood, irrespective of the complexity of the matrix $G(x)$,
from the fact that it always satisfies a first order differential equation~\eqref{eq:Abel}.
This has interesting consequences.
Quite in general, let us consider the homogeneous system Eq.~\eqref{eq:syshom} and
write it in components as
\be
\frac{d}{dx} f_i(s) = B_{ij}(x)\, f_j(x)\,.\label{eq:syscomp}
\ee
 Note that we have written the system for 
a generic set of functions $f_i(x)$, since all considerations made here
can be applied to any other $n \times n$ system of differential equations
and are not limited to the particular set of master integrals $\I_{i}(x)$ chosen above.
Let us limit ourselves to the diagonal part of the system, which reads
\be
\frac{d}{dx} f_i(x) = B_{ii}(x)\, f_i(x)\,.\label{eq:sysdiag}
\ee
It is straightforward to solve these equations by quadrature as
\be
f_{i}(x) = \exp{\left( \int_{x_0}^x \, dt\, B_{ii}(t) \right)}\,,
\ee
where $x_0$ is an irrelevant boundary condition.
With this we can define a new basis of master integrals 
\be
f_i(x) = \exp{\left( \int_{x_0}^x \, dt\, B_{ii}(t) \right)}\; \widetilde{f}_i(x)
\ee
which fulfil a new system of differential equations 
\be
\frac{d}{dx} \widetilde{f}_i(x) = \widetilde{B}_{ij}(x)\, \widetilde{f}_j(x)\,, \label{eq:systracefree}
\ee
whose matrix $\widetilde{B}(x)$ is now  traceless $\Tr \left( \widetilde{B}(x) \right) = 0$\, by construction.
Due to Abel's identity~\eqref{eq:Abel}, the Wronskian $W(x)$ of the new system becomes now particularly
simple
\be
\frac{d}{dx} W(x) = 0 \qquad \to \qquad W(x) = {\rm const}\,.
\ee
This implies in turn that, irrespective of the complexity of the functions
appearing in the solution matrix $G(x)$~\eqref{eq:Gmat}, there always exists
(at least) one  trivial combination of them such that
\be
\det{\left( G(x) \right)} = W(x) = {\rm const}\,.
\ee
In the well known case of elliptic integrals, which satisfy second-order
 differential equations, this relation reduces to the 
so-called Legendre relation.\newline

The determination of the entries of the matrix~\eqref{eq:Gmat} is in general very 
non-trivial. We will show how to do this using the information coming
from the maximal cut in the next section, as it was first suggested in~\cite{Primo:2016ebd}.
Assuming that the matrix $G(x)$ is known, one can define a new 
basis of master integrals $\M_{i}$
\begin{align}
\left(\begin{matrix} \I_{1}(x) \\ \I_{2}(x)\\ \I_{3}(x) \end{matrix} \right)=G(x)\left(\begin{matrix} \M_{1}(x) \\ \M_{2}(x)\\ \M_{3}(x) \end{matrix} \right),
\label{eq:rotatedM}
\end{align}
which by construction fulfil the system
\begin{align}
\frac{d}{dx} \left(\begin{matrix}\M_{1}(x) \\ \M_{2}(x) \\ \M_{3}(x) \end{matrix} \right) =&
\,\e \,G^{-1}(x)D(x)G(x) \left(\begin{matrix}\M_{1}(x) \\ \M_{2}(x) \\ \M_{3}(x)\end{matrix} \right)+G^{-1}(x)\left(\begin{matrix}0 \\ 0\\-\frac{1}{2 (4 x-1)}\end{matrix} \right)\,.
\label{eq:syscan}
\end{align}

Once the system is in form~\eqref{eq:syscan}, its solution order by order in $\e$
reduces, at least in principle, to a quadrature. 
In particular, the matrix $G^{-1}(x)D(x)G(x)$ (combined with the
integral over the inhomogeneous term in~\eqref{eq:syscan})
contains all information concerning the class of functions needed to iterate the solution 
at every order in $\e$.
The problem becomes then
that of classifying these functions
and of understanding their analytic and algebraic properties.
Indeed, for a general problem it will not be possible to solve these integrals in terms of known 
special functions. In our case, nevertheless, we managed to write the matrix $G(x)$ for the three-loop banana graph 
in terms of relatively simple products of elliptic integrals and it would be interesting to investigate whether a generalization
of the elliptic polylogarithms introduced to represent the two-loop massive sunrise
could be used also in this case, as the calculation in~\cite{Bloch:2014qca} seems to 
imply.

\section{The maximal cut and the homogeneous solution}
\label{sec:maxcuthomsol} \setcounter{equation}{0} 
\numberwithin{equation}{section} 

Our goal is now to solve the homogeneous system~\eqref{eq:syshom}
and determine analytically the form of the matrix $G(x)$.
A $3 \times 3$ coupled system can always be rephrased as
a third-order differential equation for any of the master integrals.
It is natural to do this for the scalar amplitude $\I_1(\e;s)$, 
whose third-order homogeneous differential equation in $d=2$ space-time dimensions 
reads
\begin{align}
\left[ \frac{d^3 }{dx^3} +\frac{3 (8 x-5)}{2 (x-1) (4 x-1)}\frac{d^2 }{dx^2}+\frac{4 x^2-2 x+1}{(x-1) x^2 (4 x-1)}\frac{d}{dx} 
+\frac{1}{x^3 (4 x-1)} \right] \I_{1}^{H}(x) = 0\,.
\label{eq:thirdeq}
\end{align}

The solution of this equation in terms of three linearly independent functions provides
the first row of the matrix $G(x)$, see Eq.~\eqref{eq:Gmat}, while the remaining two
rows can by obtained by differentiation with respect to $x$ 
\begin{align}
\label{eq:dx}
\I_{2}^H(x)=&\frac{1}{4}\left[x\frac{\partial}{\partial x}-1\right]\I_{1H}(x)\,,\\
\I_{3}^H(x)=&\frac{1}{12}\left[x^2(1-x)\frac{\partial^2}{\partial x^2}-(1+x)x\frac{\partial}{\partial x}+1\right]\I_{1H}(x).
\label{eq:d2x}
\end{align}

Solving Eq.~\eqref{eq:thirdeq} is non-trivial. 
This differential equation has been studied long ago in the context of the calculation
of cubic lattice Green functions~\cite{Joyce1}, where it was solved in terms of products of two elliptic integrals.
In order to re-derive this result 
we will use the information coming from the maximal cut. 
As it was shown in~\cite{Primo:2016ebd}, given a set of master integrals, their 
maximal cut singles out \textsl{by construction} the homogeneous part of the differential equations.
We use the notation $\Cut(\I_j(x))$ 
for the maximal cut of $\I_j(\e;s)$ evaluated in $\e = 0$
and obtain at once
\begin{align} 
\frac{d}{dx} \left(\begin{matrix} \Cut(\I_{1}(x)) \\ \Cut(\I_{2}(x))\\  \Cut(\I_{3}(x)) \end{matrix} \right) =&
B(x) \left(\begin{matrix} \Cut(\I_{1}(x)) \\ \Cut(\I_{2}(x))\\  \Cut(\I_{3}(x)) \end{matrix} \right) \, ,
 \label{eq:syscut}
\end{align}
or equivalently
\begin{align}
\left[ \frac{d^3 }{dx^3} +\frac{3 (8 x-5)}{2 (x-1) (4 x-1)}\frac{d^2 }{dx^2}+\frac{4 x^2-2 x+1}{(x-1) x^2 (4 x-1)}\frac{d}{dx} 
+\frac{1}{x^3 (4 x-1)} \right]\Cut(\I_{1}(x)) = 0\,.
\label{eq:thirdeqcut}
\end{align}
Since Eq.~\eqref{eq:thirdeqcut} admits three independent solutions, 
and all of them are required in order to construct the solution of the system,
the really interesting question becomes how one can obtain all of them 
from the maximal cut of $\I_1(x)$ only. As we showed in Section~\ref{sec:sunrise} for the simpler
case of the two-loop sunrise graph, this can be done by integrating the maximal cut along
the independent contours that do not cross any branch cuts of the integrand. Let us see how this works
for the present case.
\newline

First of all, given the definition of the banana graph in Eq.~\eqref{eq:topo},
we note that the scalar amplitude in $d=2$ dimensions is finite and can be written as 
an integral over two one-loop bubbles
\begin{align}
\I_1(x) &= \int 
\frac{\D^2 k_1 \D^2 k_2 \D^2 k_3 }{(k_1^2-m^2)((k_1-k_3)^2-m^2)(k_2^2-m^2)((k_2-(k_3+p))^2-m^2)}
\nonumber \\
&=\int \D^2 k_3 \int 
\frac{\D^2 k_1}{(k_1^2-m^2)((k_1-k_3)^2-m^2)} \int \frac{\D^2 k_2}{(k_2^2-m^2)((k_2-(k_3+p))^2-m^2)}
\end{align}
In this way, the computation of the maximal cut for the three-loop banana graph can be 
greatly simplified. 
We start with the massive one-loop bubble in $d=2$ dimensions
\begin{align}
\Bub(q^2) = \int \frac{\D^2 k}{(k^2-m^2)((k-q)^2-m^2)}\,. \label{eq:bub}
\end{align}
Its maximal cut can be easily computed to be, up to an irrelevant overall 
normalization constant,
\begin{align}
\Cut(\Bub(q^2)) = \frac{1}{\sqrt{q^2(q^2- 4 m^2)}}\,. \label{eq:cutbub}
\end{align}
Using Eq.~\eqref{eq:cutbub} the maximal cut for the banana graph can then be written as
\begin{align}
\Cut(\I_1(x)) = \oint_\C  \, 
\frac{\D^2 k_3}{\sqrt{k_3^2(k_3^2-4\, m^2)} \sqrt{(k_3+p)^2\left( (k_3+p)^2 - 4 m^2 \right)}}\,,
\end{align}
where we will specify the different choices for the contour $\C$ later on.
In $d=2$ a way to simplify this integral is to 
parametrize the loop momentum and the external one in terms of two massless
momenta $p_1^\mu$ and $p_2^\mu$ as follows
\begin{equation}
p^\mu = p_1^\mu + p_2^\mu\,, \qquad \mbox{with} \qquad p_1^2 = p_2^2 = 0\, \qquad \mbox{and} \qquad 
k_3^\mu = a \,p_1^\mu + b \,p_2^\mu\,.
\end{equation}
With this parametrization we get
\begin{align}
\oint_\C \D^2 k_3 = \frac{s}{2} \oint_{\C'} da\, db\,, \qquad \mbox{with} \quad k_3^2 = a\,b\,s  \quad \mbox{and} \quad
(k_3+p)^2 = (a+1)(b+1)\, s\,
\end{align}
and where $\C'$ represents now a still unspecified two-dimensional contour in the
four-dimensional complex hyperplane spanned
by the variables $a,b$.
In this way the integral becomes, up to a multiplicative constant
\begin{align}
\Cut(\I_1(x)) &= 
x
 \oint _{\C'}
\frac{da\, db}{\sqrt{a\,b\,(a\,b - x)} \sqrt{(a+1)(b+1) \left( (a+1)(b+1) - x \right)}} =
x\, \oint _{\C'}
\frac{da\, db}{\sqrt{R(a,b,x)}}\,, \label{eq:cutbanana}
\end{align}
where we used $x = 4\,m^2/s$ and introduced the polynomial
\be
R(a,b,x) = a\,b\,(a\,b - x)(a+1)(b+1) \left( (a+1)(b+1) - x \right)\,.
\ee

Determining explicitly which integration contours provide the 
independent solutions is a mathematically interesting problem, related to 
the dimension of the so-called cohomology group associated to the
variety described by the curve $R(a,b,c)=0$\,.
Instead of embarking into complicated mathematical considerations that go beyond 
the scope of this paper, we can try and see what we can say using simple mathematics.
Let us consider integral~\eqref{eq:cutbanana}. If we allow $a$ and $b$ to assume complex values,
we are effectively integrating on a complex two-dimensional hypersurface embedded in the 
four-dimensional complex space spanned by $a$ and $b$. There are two fundamental questions
we should answer. The first is, indeed, which contours provide a solution of the differential 
equation~\eqref{eq:thirdeq}. Once we have determined all possible contours, the second question
is which ones are linearly independent and provide therefore independent solutions.
The answer to the first question is quite general, as we claimed in Section~\ref{sec:sunrise}.
Any complex contour that does not cross any branch cuts will do the job. 
The second question, instead, has to do with recognizing which of the
allowed contours are linearly independent from each other.

It is of course very difficult to picture a two-dimensional surface embedded in a
four-dimensional space. Instead, it is useful to focus on the real two-dimensional plane
spanned by the real coordinates associated to $(a,b)$. It is trivial to see that 
the square-root in Eq.~\eqref{eq:cutbanana} has zeros for
$a=0$, $b=0$, $a=-1$, $b=-1$, $a=x/b$ and $a=x/(b+1)-1$.
We draw these sets of points in Figure~\ref{fig:contours} as continuous lines of different colors.
Similarly to the simpler case of the two-loop sunrise studied in Section~\ref{sec:sunrise},
the square root in Eq.~\eqref{eq:cutbanana} can develop a branch cut every 
time that one crosses one of these lines.
Now it is easy to convince oneself that a two-dimensional complex contour that does not cross any
branch cuts of the integrand cannot cross in particular any of the lines in Figure~\ref{fig:contours}.
We can then imagine to shrink these contours to different two-dimensional
regions in the \textsl{real plane} bounded by the branches drawn in Figure~\ref{fig:contours}.
Integrating the maximal cut~\eqref{eq:cutbanana} in any of these regions
will provide us with a viable solution of~\eqref{eq:thirdeq}.

\begin{figure}[ht!]
\begin{center}
\includegraphics[width=10cm, height=9cm]{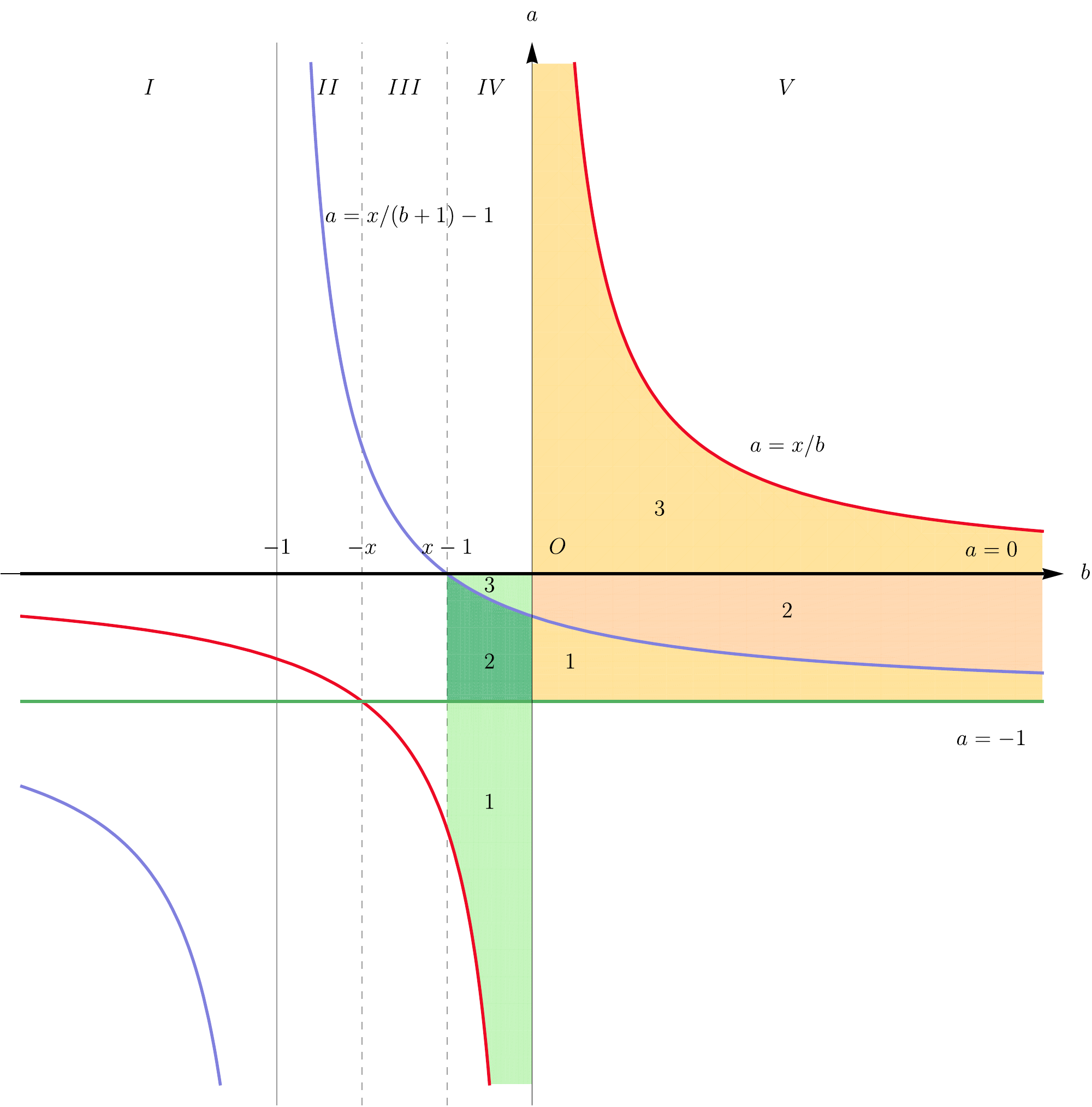}
\caption{The lines represent the set of points where the argument of the 
square root in Eq.~\eqref{eq:cutbanana}  
changes sign.} \label{fig:contours}
\end{center}
\end{figure}

Now that we know which contours are allowed, we should determine which ones of them are
independent and provide us therefore with the independent solutions of~\eqref{eq:thirdeq}.
This analysis is simplified  
by considering integral~\eqref{eq:cutbanana} as an iterated integral over two one-dimensional contours in
$da$ and $db$\footnote{Notice that the starting integral is symmetric in $a$ and $b$, so the
analysis that follows can be equally well repeated inverting the role of the two variables.}

\begin{align}
\Cut(\I_1(x)) = 
x
 \oint_{\C_b} \frac{db}{b(b+1)} \oint_{\C_a} 
\frac{da}{\sqrt{(a-a_1) \,(a-a_2)\, 
\left(a - a_3 \right) \left( a -a_4\right)}}\,, \label{eq:iterated}
\end{align}
where we introduced the abbreviations
\begin{equation}
a_1 = -1\,, \qquad a_2 = \frac{x}{b+1} -1\,, \qquad a_3 = 0\,, \qquad  a_4 =  \frac{x}{b}\,.
\label{eq:roots}
\end{equation}
From Eq.~\eqref{eq:iterated} we see that the integral in $da$ is an integral over a square root
of a quartic polynomial, which defines an elliptic curve. 
We studied this integral in section~\ref{sec:sunrise}.
That analysis implies that for every fixed value of $x$ and $b$
(which in turn determine an ordering of the roots~\eqref{eq:roots}) there are two independent contours
$\C_a$ which correspond to the two periods of the elliptic curve, see Eqs~\eqref{eq:contell}.
Without any loss of generality, we assume for the time being  $1/2 < x <1$, i.e. $4 m^2 < s < 8 m^2$.
This choice is required for the analysis below, but the result obtained is of course
independent of it and can be analytically continued to any other value of $s$.

Given this choice, we start by dividing the $(a,b)$-plane into five different regions
depending on the value of the variable $b$
\be
{\rm I}:  b\in (-\infty,-1)\,, \quad {\rm II}:  b\in (-1,-x)
\,, \quad {\rm III}:  b \in (-x, x-1) \,, \quad {\rm IV}:  b \in ( x-1, 0 )\,, 
\quad {\rm V}:  b\in (0, \infty)\,.
\label{eq:regions}
\ee
This is useful since in each region the branching points $a_j$ given in Eq.~\eqref{eq:roots} 
are ordered differently and it allows us to define \textsl{real building blocks} to construct our solutions.
This is done as follows.
First of all, for each of the regions~\eqref{eq:regions},
we can define two independent functions from the two
corresponding contours $\C_a$ identified in~\eqref{eq:contell}, and we get a total of $10$ possibilities
\begin{align}
f_1^{\rm I}(x) &= x\,\int_{-\infty}^{-1} db \int_{x/(b+1)-1}^{-1}  \frac{da}{\sqrt{-R(a,b,x)}} \,,
\qquad
f_2^{\rm I}(x) = x\,\int_{-\infty}^{-1} db \int_{-1}^{x/b}  \frac{da}{\sqrt{R(a,b,x)}}\,, 
\end{align}

\begin{align}
f_1^{\rm II}(x) &= x\,\int_{-1}^{-x} db \int_{-1}^{x/b} da \frac{da}{\sqrt{-R(a,b,x)}} \,,
\qquad
f_2^{\rm II}(x) = x\,\int_{-1}^{-x} db \int_{x/b}^{0} \frac{da}{\sqrt{R(a,b,x)}}\,, 
\end{align}

\begin{align}
f_1^{\rm III}(x) &= x\,\int_{-x}^{x-1} db \int_{x/b}^{-1}  \frac{da}{\sqrt{-R(a,b,x)}} \,,
\qquad
f_2^{\rm III}(x) = x\,\int_{-x}^{x-1} db \int_{-1}^{0}  \frac{da}{\sqrt{R(a,b,x)}}\,, 
\end{align}

\begin{align}
f_1^{\rm IV}(x) &= x\,\int_{x-1}^{0} db \int_{x/b}^{-1}  \frac{da}{\sqrt{-R(a,b,x)}} \,,
\qquad
f_2^{\rm IV}(x) = x\,\int_{x-1}^{0} db \int_{-1}^{x/(b+1)-1} \frac{da}{\sqrt{R(a,b,x)}}\,, \label{eq:funIV}
\end{align}

\begin{align}
f_1^{\rm V}(x) &= x\,\int_{0}^{\infty} db \int_{-1}^{x/(b+1)-1} \frac{da}{\sqrt{-R(a,b,x)}} \,,
\qquad
f_2^{\rm V}(x) = x\,\int_{0}^{\infty} db \int_{x/(b+1)-1}^{0} \frac{da}{\sqrt{R(a,b,x)}}\,, \label{eq:funV}
\end{align}
where the sign in the square-root is chosen to deal always with real functions.
It is very important to realize that these integral representations are not unique.
Indeed, as we discussed at length in Section~\ref{sec:sunrise}, there are different
equivalent representations for the integrals in $da$, see~\eqref{eq:contell}. 
This will turn out to be crucial 
for the considerations below.

We should notice that many of these functions
are related to each other by simple symmetry transformations.
First of all, the integrand in Eq.~\eqref{eq:cutbanana} is symmetric under $\{a \leftrightarrow b\}$
and $\{ a \to -a-1,\; b \to -b-1\}$ (and of course under a combination of the two symmetries).
It is immediate to see that under $\{ a \to -a-1,\; b \to -b-1\}$ 
region {\rm I} and {\rm V} and region {\rm II} and {\rm IV} 
are related to each other and we find
\begin{align}
f_1^{\rm I}(x) = f_1^{\rm V}(x)\,, \quad f_2^{\rm I}(x) =  f_2^{\rm V}(x)\,, \quad
f_1^{\rm II}(x) = f_1^{\rm IV}(x)\,, \quad f_2^{\rm II}(x) = f_2^{\rm IV}(x)\,. \label{eq:relcont1}
\end{align}
The effect of the symmetry under $\{a \leftrightarrow b\}$ is less immediate.
Let us take for example $f_1^{\rm V}(x)$. Using $R(a,b,x) = R(b,a,x)$, we rename 
$a$ and $b$ and  exchange the order of integration obtaining
\begin{align}
f_1^{\rm V}(x) &= x\,\int_{0}^{\infty} da \int_{-1}^{x/(a+1)-1} \frac{db}{\sqrt{-R(a,b,x)}} =
x\, \int_{-1}^{x-1} db \int_0^{x/(b+1)-1} \frac{da}{\sqrt{-R(a,b,x)}}\nonumber \\
&=x\, \int_{-1}^{-x} db \int_0^{x/(b+1)-1} \frac{da}{\sqrt{-R(a,b,x)}}
+ x\, \int_{-x}^{x-1} db \int_0^{x/(b+1)-1} \frac{da}{\sqrt{-R(a,b,x)}} \nonumber \\
&= f_1^{\rm III}(x) + f_1^{\rm IV}(x) \,, \label{eq:relcont2}
\end{align}
where in the last step we used~\eqref{eq:relcont1}. With a similar  calculation 
one can show that
\be
f_2^{\rm V}(x) = \frac{1}{2} f_2^{\rm III}(x) + f_2^{\rm IV}(x)\,, \label{eq:relcont3}
\ee
leaving in this way four independent functions, which we can choose to be
\be
f_1^{\rm V}(x)\,, \quad f_2^{\rm V}(x) \,, \quad f_1^{\rm IV}(x) \,, \quad f_2^{\rm IV}(x) \,.
\label{eq:indfunc}
\ee

The four functions~\eqref{eq:indfunc} are \textsl{not all solutions} of the third-order differential
equation. We should remember, in fact, that a solution is obtained when we integrate Eq.~\eqref{eq:cutbanana} in a region of the $(a,b)$-plane bounded by the branch cuts.
It is easy to see from Figure~\ref{fig:contours} and from the definition of the 
functions~(\ref{eq:funIV}, \ref{eq:funV}) that  while
$f_1^{\rm V}(x)$, $f_2^{\rm V}(x)$ indeed fulfil this requirement, $f_1^{\rm IV}(x)$ and
$f_2^{\rm IV}(x)$ apparently do not.
This could constitute a problem since, in order to solve a third-order differential equation
we need three independent solutions, while this argument seems to suggest we can find
only two.
Indeed, there is a subtlety. To understand it, let us look more closely at Figure~\ref{fig:contours}.
There, the integrals corresponding to these four functions are depicted as shaded areas
of different colors.
One should remember that, for any given region in $b$, different choices of integration
boundaries in $a$ produce equivalent results, as summarized in~\eqref{eq:contell}.
In particular, focussing on regions $\rm IV$ and $\rm V$, we see immediately that 
in both cases the integrals corresponding to the areas $1$ and $3$ are actually identical.
In the language of the functions defined above, this means for example that
\be
f_1^{\rm IV}(x) = x\,\int_{x-1}^{0} db \int_{x/b}^{-1}  \frac{da}{\sqrt{-R(a,b,x)}} 
= x\,\int_{x-1}^{0} db \int_{x/(b+1)-1}^{0}  \frac{da}{\sqrt{-R(a,b,x)}} \,, \label{eq:secfIV}
\ee
and the identity follows from Eq.~\eqref{eq:contell}, which specialized in this case becomes simply
$$ \int_{x/b}^{-1}  \frac{da}{\sqrt{-R(a,b,x)}}  = \int_{x/(b+1)-1}^{0}  \frac{da}{\sqrt{-R(a,b,x)}}\,.$$

As it is clear from the figure, 
the second integral representation in~\eqref{eq:secfIV}  is now integrated in a region bounded
by the branch cuts, which means that $f_1^{\rm IV}(x)$ actually also fulfills
our requirements and it is therefore a solution of the third-order differential equations.
The same cannot be said $f_2^{\rm IV}(x)$, which cannot be rewritten as
an integral over a region of plane bounded by branch cuts only.
The complete set of three independent solutions is therefore given by
$f_1^{\rm V}(x)$, $f_2^{\rm V}(x)$ and $f_1^{\rm IV}(x)$.

Interestingly, if we want to build a viable solution of the equation using the function 
$f_2^{\rm IV}(x)$, we should consider the combination
\be
f(x) = f_2^{\rm II}(x) + f_2^{\rm III}(x) + f_2^{\rm IV}(x).
\ee
As this function is defined by integrating in a region delimited by the branch cuts of the root,  
it must be a solution of the third-order differential equation. On the other hand, since we have
already determined the three independent solutions, it must be possible to write it as 
a linear combination of $f_1^{\rm V}(x)$, $f_2^{\rm V}(x)$ and $f_1^{\rm IV}(x)$.
Indeed, using~\eqref{eq:relcont1},~\eqref{eq:relcont2} and~\eqref{eq:relcont3} one finds
\be
f(x) = f_2^{\rm III}(x) +2\, f_2^{\rm IV}(x) = 2\, f_2^{\rm V}(x),
\ee
showing that the solution is indeed not linearly independent.
Of course until now we have not proved directly that the functions determined above actually 
solve the third order differential equation. We will do it later on once we have found a
more convenient representation for them.

\subsection{A basis of unit leading singularity}
Before embarking in the explicit evaluation of the integrals above, let us pause to 
interpret our results.
In the previous section we have showed that the maximal cut of the three-loop
banana graph can be evaluated on three independent contours which do not cross
branch cuts of the integrand and we claimed that the latter provide the three independent
solutions of its differential equation.
This provided us with all ingredients to build the complete homogeneous solutions of the
original $3 \times 3$ system~\eqref{eq:syshom}, as a matrix $G(x)$~\eqref{eq:Gmat}.
As we showed in Eq.~\eqref{eq:rotatedM}, the matrix $G(x)$ allows us to rotate
the original basis of master integrals $\I_1(x)$, $\I_2(x)$ and $\I_3(x)$, onto a new basis 
$M_1(x)$, $M_2(x)$ and $M_3(x)$ 
which satisfies differential equations almost entirely $\epsilon$-factorized.

The new basis defined in this way, presents a remarkable (but obvious) property. 
It's maximal cut, computed along the three independent integration contours, is unity
(or zero) by definition. This can be easily seen as follows. Inverting Eq.~\eqref{eq:rotatedM} we have

\be
\left( \begin{array}{c}  M_1(x) \\ M_2(x) \\ M_3(x) \end{array}\right) = 
\frac{1}{W(x)} \left(
\begin{array}{ccc}
 I_{3} J_{2}-I_{2} J_{3} & I_{1} J_{3}-I_{3} J_{1} & I_{2} J_{1}-I_{1} J_{2} \\
 H_{3} I_{2}-H_{2} I_{3} & H_{1} I_{3}-H_{3} I_{1} & H_{2} I_{1}-H_{1} I_{2} \\
 H_{2} J_{3}-H_{3} J_{2} & H_{3} J_{1}-H_{1} J_{3} & H_{1} J_{2}-H_{2} J_{1} \\
\end{array}
\right) \left( \begin{array}{c}  \I_1(x) \\ \I_2(x) \\ \I_3(x) \end{array}\right)\,. \label{eq:nb}
\ee
Now, from the results of the previous section, we can identify 
\begin{align}
&H_1(x) = \Cut_{\C_1}(\I_1(x)) = f_1^{\rm IV}(x)\,, \nonumber \\
&J_1(x) = \Cut_{\C_2}(\I_1(x)) = f_1^{\rm V}(x) \,,\nonumber \\
&I_1(x) =\Cut_{\C_3}(\I_1(x)) = f_2^{\rm V}(x) \label{eq:cuts1}
\end{align}
where we used the notation
$\Cut_\C(\I_j(x))$ for the maximal cut of $\I_j(x)$ computed along the contour $\C$,
which gives rise to the integration over the corresponding regions discussed above.
The remaining functions can be obtained by differentiating the ones above as in~(\ref{eq:dx}, \ref{eq:d2x}),
or alternatively by computing the maximal cuts of the other two master integrals
along the same three contours
\begin{align}
&H_2(x) = \Cut_{\C_1}(\I_2(x))\,,\quad J_2(x) = \Cut_{\C_2}(\I_2(x))  \,,\quad I_2(x) =\Cut_{\C_3}(\I_2(x)) 
\nonumber \\
&H_3(x) = \Cut_{\C_1}(\I_3(x)) \,, \quad
J_3(x) = \Cut_{\C_2}(\I_3(x))  \,,\quad I_3(x) =\Cut_{\C_3}(\I_3(x))\,. \label{eq:cuts23}
\end{align}

Using now Eq.~(\ref{eq:nb}, \ref{eq:cuts1}, \ref{eq:cuts23}) we find at once
\begin{align}
\Cut_{\C_1}(M_1(x)) = 1\,, \quad \Cut_{\C_2}(M_1(x)) = 0\,,\quad \Cut_{\C_3}(M_1(x)) = 0\,, \nonumber \\
\Cut_{\C_1}(M_2(x)) = 0\,, \quad \Cut_{\C_2}(M_2(x)) = 1\,,\quad \Cut_{\C_3}(M_2(x)) = 0\,, \nonumber \\
\Cut_{\C_1}(M_3(x)) = 0\,, \quad \Cut_{\C_2}(M_3(x)) = 0\,,\quad \Cut_{\C_3}(M_3(x)) = 1\,. \label{eq:unitcut}
\end{align}

This result is important. We can imagine, in fact, to associate to any 
family of master integrals which fulfil a set of $n$ (in our case 3) coupled differential
equations, a $n \times n$ matrix whose entries are given by the maximal cut of the integrals
evaluated along the $n$ independent integration contours.
For a generic basis of master integrals, this is by definition the matrix $G(x)$ which solves the homogeneous
system of differential equations.
For the rotated basis $M_j(x)$, Eq.~\eqref{eq:unitcut} shows that remarkably this ``matrix of maximal cuts''
reduces to the identity matrix. 

With this, we can therefore imagine a way to generalize the idea originally presented in~\cite{Henn:2013pwa},
where it was claimed that master integrals with unit leading singularity are expected to fulfil
canonical differential equations.
Suppose we are considering a basis of $n$ master integrals which fulfil $n$ coupled differential equations.
In this case, a basis which fulfils $\epsilon$-factorized differential equations must have
unit leading singularity in the sense above, i.e. the matrix which contains as entries the maximal
cut of the master integrals evaluated along all independent integration contours must be equal
to the identity matrix.
Indeed, it becomes clear that, from a practical point of view, 
there is no real difference between finding a basis of unit leading singularity and actually solving the
homogeneous system of differential equations.

\subsection{The homogeneous solutions as product of elliptic integrals}
We go back now to the explicit form of the homogeneous solution for the three-loop banana graph.
The analysis above has allowed us to determine the three independent solutions in form of 
two-fold integral representations. We might now ask ourselves if these integrals can be
performed in terms of known functions. The answer is indeed affirmative.

First of all, in order to proceed, it is useful to perform explicitly the integration in $da$ 
and obtain a one-fold integral representation for the solutions.
Using the change of variables given in~\eqref{eq:chvar} on the four function~\eqref{eq:indfunc} and after a bit of algebra 
we find respectively
\begin{align}
f_1^{\rm IV}(x) &= 2 x 
\int_{x-1}^0 \frac{db}{\sqrt{b(b+1)}\sqrt{b(b+1)+x}} 
\EK\left( 1-\frac{x^2}{b(b+1)+x} \right) \,, \label{eq:int1} \\
f_1^{\rm V}(x) &= 2x
\int_0^\infty \frac{db}{\sqrt{b(b+1)}\sqrt{b(b+1)+x}} 
\EK\left( \frac{x^2}{b(b+1)+x} \right)\,,  \label{eq:int2}\\
f_2^{\rm V}(x)
&=
2x
\int_0^\infty \frac{db}{\sqrt{b(b+1)}\sqrt{b(b+1)+x}} 
\EK\left( 1-\frac{x^2}{b(b+1)+x} \right)\,.  \label{eq:int3}
\end{align}
The integral representations (\ref{eq:int1}, \ref{eq:int2}, \ref{eq:int3}) are already more convenient
for numerical integration and analytical manipulations. Nevertheless we can do better and
compute the three integrals explicitly in terms of products of elliptic integrals of the first kind only.
The manipulations are non-trivial and we found easier to show how this is done on the last two integrals,
$f_1^{\rm V}(x)$ and $f_2^{\rm V}(x)$. Similar manipulations should be possible also for the 
first functions,
$f_1^{\rm IV}(x)$, but as it will become clear, we will not need to perform them directly.
Let us then consider Eqs. (\ref{eq:int1}, \ref{eq:int2}) and perform the change of variables
\be
b(b+1) = y^2\,, \qquad \int_0^\infty db \to \int_0^\infty \frac{2y}{\sqrt{1+4 y^2}} dy
\ee
such that the two integrals become
\begin{align}
f_1^{\rm V}(x) 
&=
2x 
\int_0^\infty \frac{dy}{\sqrt{ ( y^2+x) \, ( y^2 + 1/4)  }} 
\EK\left( \frac{x^2}{y^2+x} \right)\,, \label{eq:cut1} \\
f_2^{\rm V}(x) 
&= 
2x \int_0^\infty \frac{dy}{\sqrt{ ( y^2+x) \, ( y^2 + 1/4)  }} 
\EK\left( \frac{y^2 + x(1 - x)}{y^2+x} \right)
\,. \label{eq:cut2}
\end{align}

Eqs. (\ref{eq:cut1},\ref{eq:cut2}) can be put in standard form
by introducing three parameters $\alpha$, $\beta$ and $\gamma$ defined such that
\be
(\alpha+\beta)^2 = x\,, \quad  (\alpha-\beta)^2 = x(1-x)\,, \quad \gamma = \frac{1}{2}\,. \label{eq:defab}
\ee
In general, for a given value of $x$, four different pairs of solutions exist to these equations,
and we can choose any of those for what follows. 
For definiteness, we pick
\be
\alpha = \frac{\sqrt{x}+ \sqrt{x(1-x)}}{2}\,,\qquad \beta = \frac{\sqrt{x}- \sqrt{x(1-x)}}{2}\,,
\label{eq:choiceab}
\ee
where we are assuming $0<x<1$.
For any given pair of solutions, the integrals read
\begin{align}
f_1^{\rm V}(x) 
&= 
2x
 \int_0^\infty \frac{dy}{\sqrt{( y^2+(\alpha+\beta)^2)( y^2 + \gamma^2)  }} 
\EK\left( \frac{2\alpha \beta}{y^2+(\alpha+\beta)^2} \right) \,, \label{eq:fincut1} \\
f_2^{\rm V}(x) 
&=  
2x
\int_0^\infty \frac{dy}{\sqrt{( y^2+(\alpha+\beta)^2)( y^2 + \gamma^2)  }} 
\EK\left( \frac{y^2 + (\alpha-\beta)^2}{y^2+(\alpha+\beta)^2} \right) \label{eq:fincut2}
\,.
\end{align}
Integrals (\ref{eq:fincut1}, \ref{eq:fincut2}) are now in standard form. In particular,
the calculation of~\eqref{eq:fincut2} is discussed in~\cite{Bailey:2008ib}, see Eq.(33) therein.
Suitable extensions of the methods described there allow us to compute also also
integral~\eqref{eq:fincut1}.
The results read
\begin{align}
\label{eq:rescut1}
f_1^{\rm V}(x) 
&= 
2x
\EK(k_{-}^2)\EK(k_{+}^2)
\,\\
\label{eq:rescut2}
f_2^{\rm V}(x) 
&=
4x
\left(\EK(k_{-}^2)\EK(1-k_{+}^2)+\EK(k_{+}^2)\EK(1-k_{-}^2)\right)
\,,
\end{align}
where we have defined
\begin{align}
k_{\pm}=\frac{\sqrt{(\gamma+\alpha)^2-\beta^2}\pm \sqrt{(\gamma-\alpha)^2-\beta^2}}{2 \gamma} \qquad 
\mbox{with} \qquad k_- = \left(\frac{\alpha}{\gamma}\right) \frac{1}{\;k_+} = \frac{2 \alpha}{\;k_+}\,. \label{eq:kpm}
\end{align}
It is easy to prove by direct calculation that Eqs.~(\ref{eq:rescut1}, \ref{eq:rescut2}) solve the third
order differential equation~\eqref{eq:thirdeq}, for every choice $\alpha,\beta$ and, in particular,
for the choice we made in~\eqref{eq:choiceab}.
Even if the results for the integrals Eqs.~(\ref{eq:fincut1}, \ref{eq:fincut2}) can be found in 
standard tables of integrals, their calculation remains in our opinion not entirely straightforward and
we report it therefore in Appendix\,\ref{ap:1} following closely the methods described in~\cite{Bailey:2008ib}.
\newline

Even if we have computed only two of the solutions, by direct inspection of 
Eqs.~(\ref{eq:rescut1}, \ref{eq:rescut2}), we can easily identify the
three independent
solutions of the third-order differential equation~\eqref{eq:thirdeq} as
the three functions
\begin{align}
\Hh_{1}(x)=&
x\,
\EK\left(k_{+}^2\right)\EK\left(k_{-}^2\right)\,,\nn
\J_{1}(x)=&
x\,
\EK\left(k_{+}^2\right)\EK\left(1-k_{-}^2\right)\,,\nn
\I_{1}(x)=&
x\,
\EK\left(1-k_{+}^2\right)\EK\left(k_{-}^2\right)\,, \label{eq:sol1}
\end{align}
with $k_\pm$ given by~\eqref{eq:kpm} together with~\eqref{eq:defab} and $x = 4m^2/s$.
These results allow us to fix completely the first row of the matrix of solutions~\eqref{eq:Gmat}.
The other two rows can then be obtained by Eqs.~\eqref{eq:dx},~\eqref{eq:d2x}.
We do not report the results here for brevity, as we will use a different and more compact representation
later on.
We can nevertheless use these solutions to compute the Wronskian and find, as expected
from Abel's formula~\eqref{eq:Abel}
\begin{equation}
W(x) =-\frac{\pi ^3 x^3}{512  \sqrt{(1-4 x)^{3}(1-x)}}\,, \label{eq:Wronskfixed1}
\end{equation}
where of course the overall normalization constant depends on the explicit normalization
choice made on Eqs.~\eqref{eq:sol1}.
Inspecting our three solutions, it is natural to wonder what would 
happen considering one further combination
\be
\K_{1}(x) = 
x\,
\EK\left(1-k_{+}^2\right)\EK\left(1-k_{-}^2\right)\,. \label{eq:sol1plus}
\ee
It is simple to prove by direct calculation that also Eq.~\eqref{eq:sol1plus} solves the differential
equation~\eqref{eq:thirdeq}. Of course, since a third-order differential equation admits only
three independent solutions, this last solution cannot be linearly independent from the previous three.
Indeed, it is easy to check numerically (for example using the PSLQ algorithm) that this is true. 
Since the four functions develop imaginary parts for $x<0$ or $x > 1/4$, 
the exact relation between the solutions depends on the value of the variable $x$ and on 
the convention picked for their analytic continuation. 
With the choice~\eqref{eq:choiceab}
and taking $0<x<\frac{1}{4}$ ($s > 16 m^2$), for which all solutions
Eqs.~(\ref{eq:sol1},\ref{eq:sol1plus}) are real valued, one finds
\be
\K_{1}(x)\big|_{0<x<\frac{1}{4}} = \frac{1}{3} \Hh_1(x)\big|_{0<x<\frac{1}{4}}\,.
\ee
\newline

One last comment is in order. We have computed explicitly two out of the three functions that we claimed
constitute the three solutions of the third-order differential equation.
We have been lucky enough to be able to extract from these two functions a representation for
\textsl{all three independent solutions} of the equation as products of elliptic integrals~\eqref{eq:sol1}.
This of course has to be considered as an accident. One might wonder what would the calculation
of the remaining function $f_1^{\rm IV}(x)$ have produced. Indeed, if it is true that it is also a solution
of the equation, and if the solutions chosen in~\eqref{eq:sol1} are really independent, it must be possible to
write $f_1^{\rm IV}(x)$ as linear combination of the latter.
Instead of performing the integration explicitly, it is simple to prove this by use of the PSLQ algorithm.
As before, the relation depends on the value of $x$.
The function $f_1^{\rm IV}(x) $ is real-valued for $1/4<x<1$, while the three solutions~\eqref{eq:sol1} are
real-valued for $0<x<1/4$.
Picking then  $1/4<x<1$ and assuming $x \to x + i\, 0^+$, one finds easily the following relation
\be
f_1^{\rm IV}(x) = 4\, \left( \Hh_1(x) + i\, \J_1(x) - i\, \I_1(x) \right)\,,
\ee
showing that, as expected, also $f_1^{\rm IV}(x)$ is a solution of the third-order
differential equation satisfied by the banana graph.

\section{The third-order differential equation as a symmetric square}
\label{sec:symmsquare} \setcounter{equation}{0} 
\numberwithin{equation}{section} 
In the previous section we have showed that different solutions of the third-order differential 
equation~\eqref{eq:thirdeq} can be found by integrating the maximal cut of the three-loop massive
banana graph along independent integration contours. 
Here we want to elucidate the relation of the solutions found
above with an alternative set of solutions found by G.S. Joyce~\cite{Joyce1}.
For simplicity, let us rewrite here the third-order homogeneous differential equation 
satisfied by the three-loop banana graph

\begin{align}
\left[ \frac{d^3 }{dx^3} +\frac{3 (8 x-5)}{2 (x-1) (4 x-1)}\frac{d^2 }{dx^2}+\frac{4 x^2-2 x+1}{(x-1) x^2 (4 x-1)}\frac{d}{dx} 
+\frac{1}{x^3 (4 x-1)} \right] \I_{1}^{H}(x) = 0\,.
\end{align}
This equation has a remarkable property, i.e. it is a so-called
\textsl{symmetric square}.
Completely in general, let $L_3(x)$ be a third-order differential operator 
\be
L_3(x) = \frac{d^3}{d^3 x} + c_2(x) \frac{d^2}{d^2 x}  + c_1(x) \frac{d}{d x} + c_0(x)\,. \label{eq:third}
\ee
The operator $L_3(x)$ is a \textsl{symmetric square} if its three independent solutions can
be written as
\be
g_1(x) = \left( f_1(x) \right)^2\,,\qquad g_2(x) = f_1(x)\, f_2(x)\,, \qquad g_3(x) = \left( f_2(x) \right)^2
\label{eq:symcomb}
\ee
where the two functions $f_1(x)$ and $f_2(x)$ are in turn solutions of a second-order differential operator
\be
L_2(x) =\frac{d^2}{d^2 x}  + a_1(x) \frac{d}{d x} + a_0(x)\,. \label{eq:second}
\ee
We have therefore
\be
L_2(x) f_1(x) = L_2(x) f_2(x) = 0\,,
\ee
and correspondingly
\be
L_3(x) g_1(x) = L_3(x) g_2(x) = L_3(x) g_3(x) = 0\,.
\ee

Testing whether a third-order differential operator is a symmetric square and building the
corresponding second-order differential operator is very simple. Starting from the coefficients
of $L_3(x)$ in~\eqref{eq:third} we can build the following two combinations
\be
\alpha_1(x) = \frac{1}{3}c_2(x)\,, \qquad \alpha_0(x) = \frac{c_1(x) - \alpha_1'(x) - 2 \alpha_1^2(x)}{4}\,,
\label{eq:test1}
\ee
where $\alpha_1'(x) = d \alpha_1(x)/dx$.
Now, if the following relation is satisfied
\be
4\, \alpha_0(x) \alpha_1(x) + 2 \, \frac{d \alpha_0(x)}{dx} = c_0(x)\,,
\ee
then the differential operator $L_3(x)$ in~\eqref{eq:third} is the symmetric square of
the corresponding second-order differential operator
\be
L_2(x) =\frac{d^2}{d^2 x}  + \alpha_1(x) \frac{d}{d x} + \alpha_0(x)\,. \label{eq:second2}
\ee

It is straightforward to check whether our third-order differential equation~\eqref{eq:thirdeq}
is a symmetric square. Starting with
\begin{align}
c_2(x) = \frac{3 (8 x-5)}{2 (x-1) (4 x-1)} \,,\qquad 
c_1(x) = \frac{4 x^2-2 x+1}{(x-1) x^2 (4 x-1)} \,, \qquad
c_0(x) = \frac{1}{x^3 (4 x-1)}\,,
\end{align}
we immediately find the corresponding coefficients
\begin{align}
\alpha_1 = \frac{8 x-5}{2 (x-1) (4 x-1)}\,, \qquad
\alpha_0 = -\frac{2 x-1}{4 (x-1) x^2 (4 x-1)}\,,
\end{align}
and indeed one can verify that
\begin{align}
4 \, \alpha_0(x) \alpha_1(x) + 2 \, \frac{d \alpha_0(x)}{dx}  = \frac{1}{x^3 (4 x-1)} = c_0(x)\,.
\end{align}
This shows that the solutions of the third-order differential equation satisfied
by the three-loop banana graph~\eqref{eq:thirdeq}
can be written as symmetric square combinations of the two solutions of the
following second-order differential equation
\be
\left[ \frac{d^2}{d^2 x}  + \frac{8 x-5}{2 (x-1) (4 x-1)} \frac{d}{d x} -\frac{2 x-1}{4 (x-1) x^2 (4 x-1)}\right] f(x) = 0\,,
\ee
which can be then solved in terms of a class of special functions called Heun functions~\cite{Joyce1}.
Interestingly enough, one can show that such Heun functions can be rewritten as a product of
elliptic integrals of suitable arguments. The result is very non-trivial and we refer to~\cite{Joyce1}
and references therein for details. 
The three solutions found there read
\begin{align}
&H_{1}(x) =\EK\left(\omega_{+}\right) \EK\left(\omega_{-}\right)\,,\nn
&J_{1}(x)  =\EK\left(\omega_{+}\right) \EK\left(1-\omega_{-}\right)\,,\nn
&I_{1}(x)   =-\frac{1}{3}\EK\left(1-\omega_{+}\right)\EK\left(1-\omega_{-}\right)\,,
\label{eq:solJoyce1}
\end{align}
where we defined
\begin{align}
\omega_{\pm}=\frac{1}{4x}\left(2x+(1-2x)\sqrt{\frac{x-1}{x}}\pm\sqrt{\frac{4x-1}{x}}\right)\,.
\end{align}
Moreover, in~\cite{Joyce1} it is shown that a relation exists between the elliptic integrals above.
If we assume $x>1$ than the functions~\eqref{eq:solJoyce1} are explicitly real and the relation reads
\be
\EK(\omega_{+}) = \left( \sqrt{ \frac{4x-1}{x}} - \sqrt{1 - \frac{1}{x}}\right) \EK(\omega_-)\,.
\ee
This representation of the solutions is somewhat more compact than what we found
above in~\eqref{eq:sol1}, but of course the two sets of solutions must be equivalent
since they solve the same third-order differential equation. Proving by algebraic manipulations 
that the two set of solutions can be written one in terms of the other is non trivial,
in particular since the exact relations among the two depends on the region in $x$ which
we consider. Indeed, every time one crosses a branching point
of the differential equation~\eqref{eq:thirdeq}, each of the three solutions gets in general
mapped to a linear combination of the same three functions.
To give an example, we consider again the region $x > 1$ where the functions~\eqref{eq:solJoyce1}
are real. The solutions found in~\eqref{eq:sol1} instead are complex and we need a prescription, which
we choose for definiteness to be $x \to x + i 0^+$. By using PSLQ one then finds
(with virtually arbitrary precision) that the following relations are satisfied
\begin{align}
&H_{1}(x) = 2\,i\, \Hh_1(x) - \J_1(x) + 3 \I_1(x)\,,\nonumber \\
&J_{1}(x) = -2\,i\, \Hh_1(x) + 3\J_1(x) - \I_1(x)\,,\nonumber \\
&I_{1}(x) =  -i\,  \J_1(x) + i\, \I_1(x)\,,
\end{align}
showing that the two sets of solutions are indeed equivalent.
The remaining two columns of the matrix of solutions $G(x)$ can then be determined 
differentiating~\eqref{eq:sol1} as in Eqs.~(\ref{eq:dx}, \ref{eq:d2x})\,.
Using these solutions one find for the Wronskian
\be
W(x) = \frac{\pi ^3 x^3}{64 \sqrt{ (4 x-1)^{3}(x-1)}}\,. \label{eq:wronskfin}
\ee
\newline

We obtained in this way two equivalent representations for the entries of the matrix of homogeneous
solutions $G(x)$ defined in~\eqref{eq:Gmat}. Still we are not quite done.
Since the differential equation~\eqref{eq:thirdeq} has four singular
points $x=0$, $x=1/4$, $x=1$ and $x=\infty$, its solutions can develop branch cuts crossing
any of these points. 
It is easy to see that the solutions built in this section are real for $x>1$ but they develop
an imaginary part whenever $x<1$. Moreover they have discontinuities in all other singular points.
On the other hand, the solutions found in~\eqref{eq:sol1} turn out to be real only for $0<x<1/4$.
In order to properly analytically continue the results for every value of $x$ 
we will need to build other solutions, similar to~\eqref{eq:solJoyce1} or~\eqref{eq:sol1} , but which are real
in the remaining regions $(-\infty,0)$, $ (1/4,1)$. Indeed, in every region the solutions can be built taking
simple linear combinations of~\eqref{eq:sol1} or~\eqref{eq:solJoyce1}. Many different combinations are possible  and we refer to Appendix~\ref{app:conthom} for details on one possible choice which we found convenient.
To indicate these different sets of solutions we introduce the notation
 $H_k^{(a,b)}(x)$, $J_k^{(a,b)}(x)$, $I_k^{(a,b)}(x)$, where the superscript 
 indicates that the corresponding solution is real for $a<x<b$. The corresponding
 matrix of solutions will then be indicated as $G^{(a,b)}(x)$. We normalize our solutions
 for all values of $x$ in such a way that the corresponding Wronskian $W^{(a,b)}(x)$
is equal to~\eqref{eq:wronskfin}, up to a possible overall factor $i$ which is required
 when the argument of the square-root becomes negative.

\section{The inhomogeneous solution}
\label{sec:inhomsol} \setcounter{equation}{0} 
\numberwithin{equation}{section} 
In this section we make use of the homogeneous solutions of the system of differential 
equations \eqref{eq:topsys2}, which have been studied in 
Sections~\ref{sec:maxcuthomsol}-\ref{sec:symmsquare} and collected in Appendix~\ref{app:conthom}, 
in order to build the inhomogeneous solution as a series expansion around $d=2$.\\
The following discussion holds for any of the kinematic regions $a<x<b$ located by the four singular 
points of the differential equations and we will keep giving as understood the superscript $(a,b)$.\\
Both basis of master integrals $\I_{i}(x)$ and $\M_{i}(x)$ are finite in $d=2$ 
and they can be Taylor-expanded as
\begin{align}
\I_{i}(x)=\sum_{k=0}^{\infty}\I_{i}^{(k)}(x)\eps^{k}\,,\qquad \M_{i}(x)=\sum_{k=0}^{\infty}\M_{i}^{(k)}(x)\eps^{k}\,\quad i=1,2,3.
\label{eq:taylor}
\end{align}
By substituting Eq.~\eqref{eq:taylor} into \eqref{eq:syscan}, we obtain a particularly simple set 
of chained first order differential  equations for the coefficients $\M_{i}^{(k)}(x)$, which reads
\begin{align}
\frac{d}{dx} \left(\begin{matrix} \M_{1}^{(0)} \\ \M_{2}^{(0)} \\ \M_{3}^{(0)}\end{matrix} \right) =&
G^{-1}(x) \left(\begin{matrix}0 \\ 0 \\ \frac{1}{2(1-4x)}\end{matrix} \right)
\label{eq:syseps0}
\end{align}
and
\begin{align}
\frac{d}{dx} \left(\begin{matrix}\M_{1}^{(n)}(x) \\ \M^{(n)}_{2}(x)\\ \M_{3}^{(n)}(x) \end{matrix} \right) =&
\,\e \,G^{-1}(x)D(x)G(x) \left(\begin{matrix}\M_{1}^{(n-1)}(x)\\ \M_{2}^{(n-1)}(x) \\ \M_{3}^{(n-1)}(x)\end{matrix} \right)\,,\quad n>0.
\label{eq:sysepsn}
\end{align}
From Eq.\eqref{eq:sysepsn} we see that the solution for the master integrals $\M_{i}(x)$ 
has a manifest iterative structure, since each coefficient $\M_{i}^{(n)}(x)$ can be simply written 
as an integral of the lower order terms  $\M_{i}^{(k-1)}(x)$, convoluted with the integration kernel $G^{-1}(x)D(x)G(x)$.
Together with the integration of the lowest order~\eqref{eq:syseps0}, this kernel specifies 
the class of functions required at every order in $\e$.

Once $\M_{i}^{(n)}(x)$ have been determined, the corresponding term of the 
$\eps$-expansion of the original master integrals $\I _{i}(x)$ can be obtained by applying 
the rotation matrix $G(x)$ back to the integrals $\M_{i}(x)$, according to the definition \eqref{eq:rotatedM},
\begin{align}
\left(\begin{matrix} \I_{1}^{(n)}(x) \\ \I_{2}^{(n)}(x)\\ \I_{3}^{(n)}(x) \end{matrix} \right)=G(x)\left(\begin{matrix} \M_{1}^{(n)}(x) \\ \M_{2}^{(n)}(x)\\ \M_{3}^{(n)}(x) \end{matrix} \right)\,.
\end{align}
In the remaining of this section, we will limit ourselves to the determination of the order zero terms $\I _{i}^{(0)}(x)$.
The latter, by definition, are not the entire story, as their calculation does not require to integrate over the
kernel $G^{-1}(x) D(x) G(x)$. While we could extend the methods described in this section in order to provide integral
representations also for the higher orders, we do not find this particularly useful and postpone this problem
to later work. A complete solution of this problem, in fact, would require to understand and classify the
properties of the functions defined by repeated integrations over the kernel above.
\\
The differential equations \eqref{eq:syseps0} can be readily solve solved by quadrature, producing
\begin{align}
 \M_{1}^{(0)}(x)=c_{1}^{(0)}+\int_{x_0}^{x}dt\frac{1}{1-4t}\mathcal{R}_{1}(t)\,,\nn
 \M_{2}^{(0)}(x)=c_{2}^{(0)}+\int_{x_0}^{x}dt\frac{1}{1-4t}\mathcal{R}_{2}(t)\,,\nn
 \M_{3}^{(0)}(x)=c_{3}^{(0)}+\int_{x_0}^{x}dt\frac{1}{1-4t}\mathcal{R}_{3}(t)\,,
 \label{Msol}
\end{align}
where the integration base-point $x_0$ can be arbitrarily chosen and the integration constants $c_{i}^{(0)}$ have to be fixed by imposing suitable boundary conditions. 
The integrands $\mathcal{R}_{i}(x)$ are combinations of products of two homogeneous solutions 
which originate from the entries of $G^{-1}(x)$ (see Eq.~\eqref{eq:Ginv}),
\begin{align}
\mathcal{R}_{1}(x) 
=& \frac{1}{2W{(x)}}\left[I_{2}(x)J_{1}(x)-I_{1}(x)J_{2}(x)\right]\,,\nn
\mathcal{R}_{2}(x)
=& \frac{1}{2W(x)}\left[H_{2}(x)I_{1}(t)-H_{1}(x)I_{2}(x)\right]\,,\nn
\mathcal{R}_{3}(x)
=& \frac{1}{2W(x)}\left[H_{1}(x)J_{2}(t)-H_{2}(x)J_{1}(x)\right]\,.
\label{rotatedM1}
\end{align}
Therefore eqs.~\eqref{Msol} and \eqref{rotatedM1} completely specify the inhomogeneous solution at order zero 
once the boundary constants $c_{i}^{(0)}$ are fixed, for instance  by imposing the regularity of the solutions at specific kinematic points.\\
We have already observed that the system \eqref{eq:topsys2}, or equivalently the third order differential equation \eqref{eq:thirdeq}, 
has regular singular points at $x=1$ and $x=\pm \infty$, which correspond, respectively, to the pseudo-thresholds $s=4m^2$ and $s=0$ 
of the equal-mass banana graph.  One can show that demanding the regularity of $\I_{i}^{(0)}(x)$ 
at such points is indeed sufficient in order to fix the three integration constants $c_{i}^{(0)}$. 
In fact, by imposing regularity directly on the system of differential equations, 
one can determine three independent linear relations, 
which must be satisfied by the master integrals on the pseudo-thresholds.
In particular, regularity at $x\to 1^{\pm}$ requires 
\begin{align}
\lim_{x\to 1^{\pm}}\left(\I_{3}^{(0)}(x)+\frac{2}{3}\I_{2}^{(0)}(x)+\frac{1}{12}\I_{1}^{(0)}(x)\right)=0\,,
\label{eq:lim1p}
\end{align}
whereas at $x\to  \pm\infty$ we find
\begin{align}
\label{eq:reg1}
\lim_{x\to \pm\infty}\left(\I_{2}^{(0)}(x)+\frac{1}{4}\I_{1}^{(0)}(x)\right)=0\,,\\
\lim_{x\to  \pm\infty}\left(\I_{3}^{(0)}(x)-\frac{1}{16}\I_{1}(x)^{(0)}\right)=\frac{1}{8}\,.
\label{eq:reg2}
\end{align}
It is worths observing that, since $x\to  \pm\infty$ corresponds to $s\to 0^{\pm}$, the two conditions 
(\ref{eq:reg1},~\ref{eq:reg2}) consistently reproduce the IBPs identities between the three-loop vacuum 
diagrams to which the master integrals are reduced in the zero-momentum limit.\\

\begin{figure}[h!]
  \centering
  \subfloat[][]{%
    \includegraphics[width=0.50\textwidth,trim=0 0 0 0.27cm]{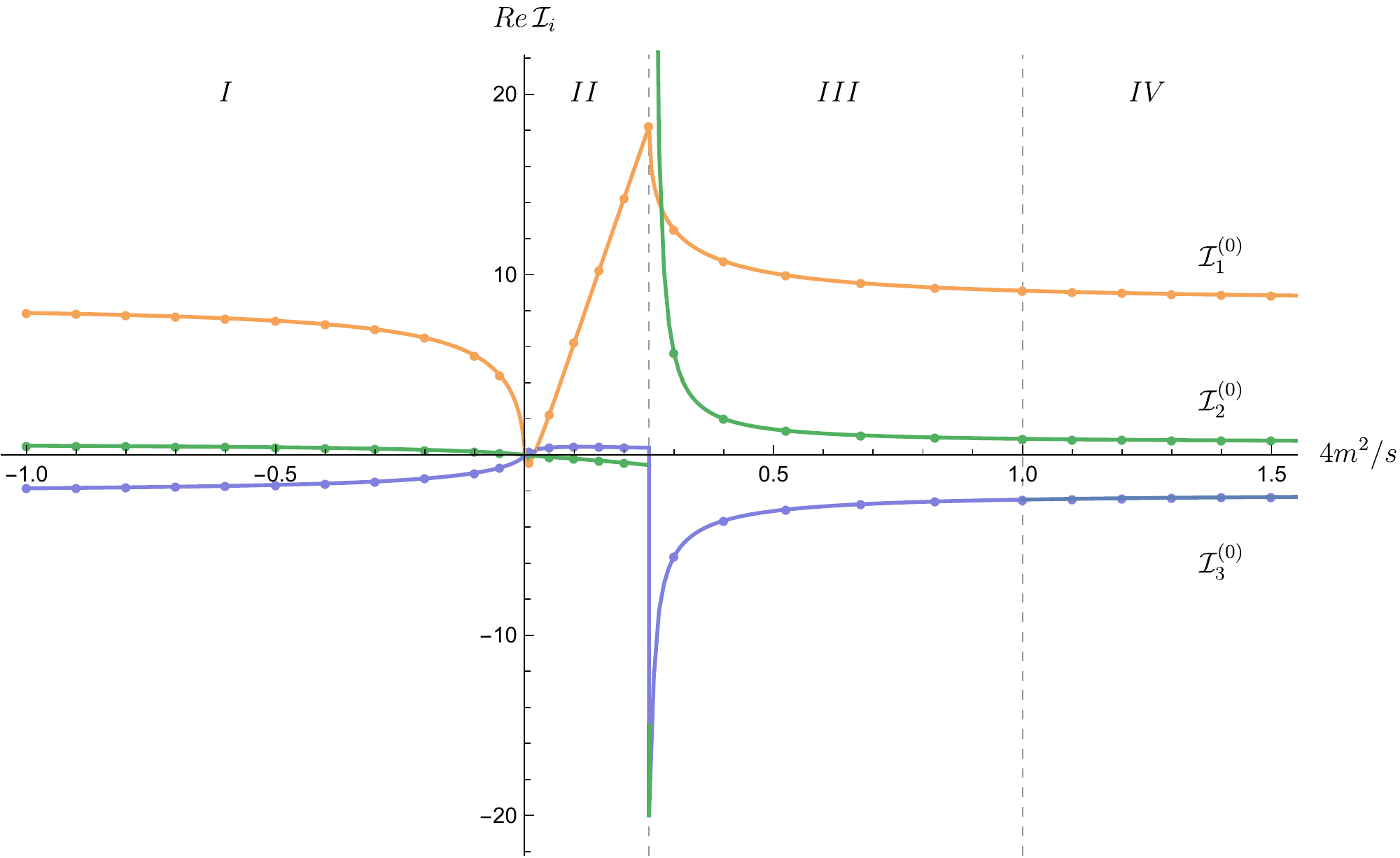}
    \label{fig:PlotRe} }
  \subfloat[][]{%
    \includegraphics[width=0.51\textwidth,trim=0 0 0 0.27cm]{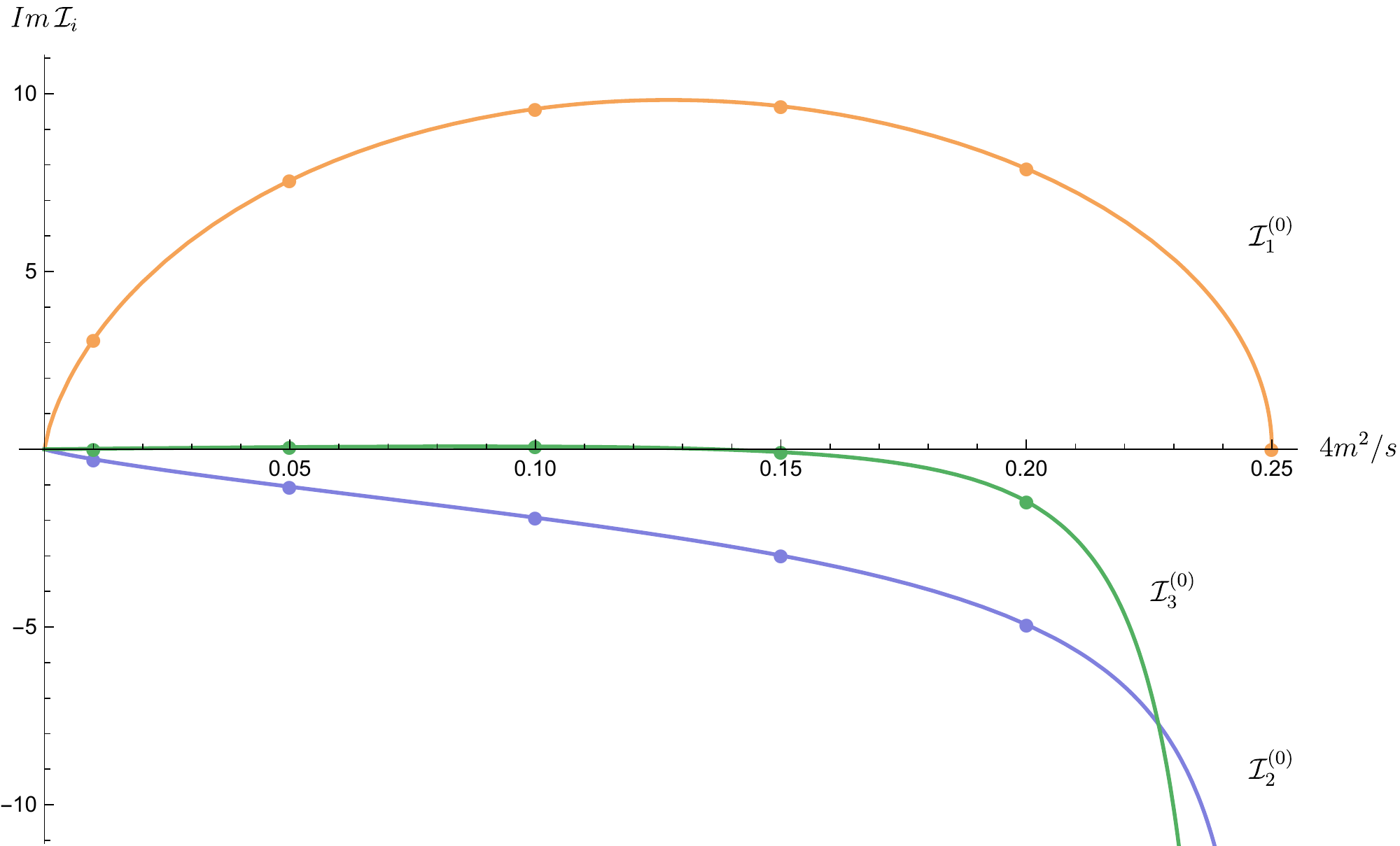}
      \label{fig:PlotIm}
  }
  \caption{Real \protect\subref{fig:PlotRe} and imaginary \protect\subref{fig:PlotIm} part of the finite term of the master integrals for the three-loop banana graph. The imaginary part is non-vanishing only in the range $0<x<1/4$, which corresponds to $s>16m^2$. The numerical evaluation of the result (solid curves) is compared to the values obtained with SecDec 3 (dots).}
  \label{fig:diags}
\end{figure}

It is particularly convenient to fix explicitly the boundary constants by working in the region $1<x<\infty$, 
since the end-points of this region corresponds exactly to the two pseudo-threshold where we impose 
the regularity conditions (\ref{eq:lim1p}-\ref{eq:reg2}). 
If we specify Eq.~\eqref{Msol} to the interval $(1,\infty)$ and apply the rotation \eqref{rotatedM1}, we get
\begin{align}
\I_{1}^{(0)}(x)=&H^{(1,\infty)}_{1}(x) \left[c_{1}^{(0)}+\int_{1}^x dt \frac{1}{1-4t}\mathcal{R}^{(1,\infty)}_{1}(t)\right]+
J^{(1,\infty)}_{1}(x)\left[ c_{2}^{(0)}+\int_{1}^xdt \frac{1}{1-4t}\mathcal{R}^{(1,\infty)}_{2}(t)\right]+\nn
&I^{(1,\infty)}_{1}(x) \left[c_{3}^{(0)}+\int_{1}^xdt \frac{1}{1-4t}\mathcal{R}^{(1,\infty)}_{3}(t)\right]
 \,,\nn
\I_{2}^{(0)}(x)=&H^{(1,\infty)}_{2}(x) \left[c_{1}^{(0)}+\int_{1}^xdt \frac{1}{1-4t}\mathcal{R}^{(1,\infty)}_{1}(t)\right]+J^{(1,\infty)}_{2}(x)\left[ c_{2}^{(0)}+\int_{1}^xdt \frac{1}{1-4t}\mathcal{R}^{(1,\infty)}_{2}(t)\right]+\nn
&I^{(1,\infty)}_{2}(x) \left[c_{3}^{(0)}+\int_{1}^xdt \frac{1}{1-4t}\mathcal{R}^{(1,\infty)}_{3}(t)\right]\,,\nn
\I_{3}^{(0)}(x)=&H^{(1,\infty)}_{3}(x) \left[c_{1}^{(0)}+\int_{1}^xdt \frac{1}{1-4t}\mathcal{R}^{(1,\infty)}_{1}(t)\right]+J^{(1,\infty)}_{3}(x)\left[ c_{2}^{(0)}+\int_{1}^xdt \frac{1}{1-4t}\mathcal{R}^{(1,\infty)}_{2}(t)\right]+\nn
&I^{(1,\infty)}_{3}(x)\left[ c_{3}^{(0)}+\int_{1}^xdt \frac{1}{1-4t}\mathcal{R}^{(1,\infty)}_{3}(t)\right] \,,
\label{eq:Isol}
\end{align}
where we have chosen as integration-base point $x_0=1$ and re-introduced the superscript $(1,\infty)$ for all quantities that require
analytic continuation.
We remark that, when applied to Eq.~\eqref{eq:Isol}, the definition \eqref{rotatedM1} of the function $\mathcal{R}^{(1,\infty)}_{i}(x)$ 
must be interpreted in terms of the homogeneous solutions $G^{(1,\infty)}(x)$, which are defined in \eqref{eq:homI}. 
Due to the choice of the integration base-point, in the $x\to 1^{+}$ limit all integrals appearing in the r.h.s of \eqref{eq:Isol} 
vanish and the master integrals become
\begin{align}
\lim_{x\to 1^{+}}\I_{i}^{(0)}(x)=&\lim_{x\to 1^{+}}\left(c_{1}^{(0)}H^{(1,\infty)}_{i}(x) +c_{2}^{(0)}J^{(1,\infty)}_{i}(x)+c_{3}^{(0)}I^{(1,\infty)}_{i}(x)\right)  \,,\quad i=1,2,3.
\label{eq:regI1}
\end{align}
The limiting behaviours of the homogenous solutions $H_{i}(x)$, $J_{i}(x)$, $I_{i}(x)$ at the two pseudo-thresholds are discussed in Appendix~\ref{app:conthom} and it is easy to verify that, when the expansions at $x\to 1^{+}$ \eqref{eq:exp1p} are plugged into Eq.~\eqref{eq:regI1}, the regularity constraint \eqref{eq:lim1p} is satisfied by demanding
\begin{align}
c_{3}^{(0)}=-3c_{1}^{(0)}\,.
\label{eq:c3}
\end{align}
In a similar way, we can impose regularity at $x\to +\infty$  by making use of the expansions \eqref{eq:expinf}. Remarkably, due to the presence of logarithmic divergences $\ln(1/x)$ in the expansion of the homogenous solutions which must cancel in the expression of the master integrals, Eq.~\eqref{eq:reg1} allows us to fix at once $c_{1}^{(0)}$ and $c_{2}^{(0)}$, which are given by
  
\begin{align}
c_{1}^{(0)}=&\frac{1}{3}\int_{1}^{\infty}dt\frac{1}{(1-4t)}\mathcal{R}^{(1,\infty)}_{3}\left(t\right)=\frac{1}{3}\int_{0}^{1}\frac{dy}{y(y-4)}\mathcal{R}^{(1,\infty)}_{3}\left(\frac{1}{y}\right)\,,\nn
c_{2}^{(0)}=&-\int_{1}^{\infty}\frac{dt}{(1-4t)}\mathcal{R}^{(1,\infty)}_{2}\left(t\right)=-\int_{0}^{1}\frac{dy}{y(y-4)}\mathcal{R}^{(1,\infty)}_{2}\left(\frac{1}{y}\right)\,.
\label{eq:c1c2}
\end{align}
In the second equalities we have simply performed the change of variable $t\to 1/y$ in order to map the integration range to $0<y<1$. As a consistency check, we have verified that these values of the integration constants are consistent also with the regularity condition for the third master, given by Eq.~\eqref{eq:reg2}.\\Although we were not able to determine an analytic expression of the boundary constants, their representation as definite integrals \eqref{eq:c1c2} allows a high-precision numerical evaluation. We get for example
 \begin{alignat}{2}
 &c_{1}^{(0)}=&&-1.2064599496517629858762117245910770452963348722...
 \nonumber \\
 &c_2^{(0)} = &&+2.5819507507087486799289938331551672385057488393...
 \end{alignat}
The representation \eqref{eq:Isol} of the master integrals, which is now fully determined, is valid for $1<x<\infty$. The expression of $\I_{i}^{(0)}$ in the other kinematic regions (and in particular for $0<x<1/4$, $s>16m^2$, where the master integrals develop an imaginary part) can be obtained by analytic continuation of Eq.~\eqref{eq:Isol}. 
The details of the analytic continuation are presented in Appendix~\ref{app:conthom}. 
As a summary of the results there discussed, let us just stress that the determination of a 
set explicitly real solutions $G^{(a,b)}(x)$ in each region and the study of their leading 
behaviour in the proximity of the singular points can be used to define, for any value of $x$, 
a representation of the solutions which involves individually real-valued integrals only and, therefore, 
allows a fast and accurate numerical evaluation. 
A plot of the numerical results obtained through our representation compared against 
the computer code SecDec 3~\cite{Borowka:2015mxa} is shown in Fig.~\ref{fig:diags}.

\section{Conclusions}
In this paper, we dealt with two important issues for the calculation of complicated multiloop Feynman integrals.
The first, which has recently received a lot of attention, has to do with the explicit construction
of a full set of homogeneous solutions of the system of couple differential equations satisfied by
a family of Feynman integrals. The second, involves the classes of functions that are required for their calculation,
 beyond the well understood multiple polylogarithms.
To address both issues in a non-trivial environment, we considered the calculation of the three-loop
massive banana graph, which is the simplest graph known to satisfy an irreducible third-order
differential equation. 
Complementing the findings reported in~\cite{Primo:2016ebd}, we showed that all independent
homogeneous solutions can be found by evaluating the maximal cut of the graph on the 
independent contours which do not cross any branch cuts of
the integrand. We showed explicitly that in the case of the three-loop massive banana graph, there are
only three possible choices of independent contours and that they indeed provide all three independent
homogeneous solutions of its third-order differential equation. Our findings are in perfect agreement
with the ones recently presented in~\cite{Bosma:2017ens}.

Given the three independent homogeneous solutions as integrals over contours, we showed
how to evaluate them explicitly in terms of products of complete elliptic integrals. The result, which
was found long ago with very different methods by Joyce~\cite{Joyce1}, is very interesting for two reasons.
On the one hand, it is the first known generalization of the two-by-two differential equations satisfied, for example,
by the two-loop sunrise graph, and which give rise to integrals over elliptic integrals.
On the other hand, while generalizing the case above, it does not require to introduce truly new functions, as the
homogeneous solutions can be still written as products of elliptic integrals only.
While this is known to be an accident of the equal-mass case, it is nevertheless a very interesting laboratory
to study the functions required to evaluate Feynman integrals which fulfil differential equations
of order higher than two.

The results described in this paper show that the maximal cut can be used to solve complicated
homogeneous differential equations, even when the number of coupled equations is higher than two.
At variance with the $2 \times 2$ case, which seems to invariably require complete elliptic integrals, 
$3 \times 3$ and
higher cases might require the introduction and classification of more complicated functions.
We look forward to applying this method to increasingly complicated cases of relevance for
high-energy physics phenomenology.

\section*{Acknowledgements}
One of the authors (L.T.) is grateful to the Mainz Institute for Theoretical Physics (MITP), where some of 
the ideas at the basis of this project were developed. 
We wish to thank in particular Johannes Henn and Stefan M\"{u}ller-Stach, whose encouragement and 
suggestions were crucial at different stages of the project.
We also acknowledge many clarifying discussions with Claude Duhr, Harald Ita, Pierpaolo Mastrolia and Ettore Remiddi.
Finally, we wish to thank David Broadhurst for interesting comments and for pointing us to the correct
literature on cubic lattice Green functions.
 A.P. wishes to thank the Institute for Theoretical Particle Physics of the Karlsruhe Institute of Technology 
 for hospitality during the development of this project.

\appendix

\section{Analytic continuation}
\label{app:conthom}
In this appendix we describe the analytic continuation of the master integrals derived in 
Section~\ref{sec:inhomsol} to arbitrary values of $x=4m^2/s$. We start from the analytic continuation 
of the homogenous solutions by first defining, for each kinematic region $a<x<b$, a set of real-valued 
homogeneous solutions $G^{(a,b)}(x)$ and then by matching their limiting behaviours 
in order to link them across the singularities of the differential equation. 
Finally we will apply these results to Eq.~\eqref{eq:Isol} and obtain the analytic continuation of the inhomogeneous solution.
\subsection{Homogenous solutions}
In Sections~\ref{sec:maxcuthomsol}-\ref{sec:symmsquare} we have obtained, through different approaches, 
two different representation of the homogenous solutions, corresponding to Eq.s~\eqref{eq:sol1} and \eqref{eq:solJoyce1}. 
Although the two representations have been shown to be completely equivalent, we decide to work with the latter, 
since it leads to more compact expression. Therefore, we  consider homogenous solutions written in 
terms of products of complete elliptic integrals of with arguments
\begin{align}
\omega_{\pm}=\frac{1}{4x}\left(2x+(1-2x)\sqrt{\frac{x-1}{x}}\pm\sqrt{\frac{4x-1}{x}}\right)\,.
\end{align}
The solutions \eqref{eq:solJoyce1} are explicitly real in the region $(1,\infty)$. In order to define a set of real solutions in the other three regions, $(-\infty,0)$, $(0,1/4)$ and $(1/4,1)$, we can make use of well-known identities between complete elliptic integrals such as
\begin{align}
\EK\left(\frac{1}{z}\right)=&\sqrt{z}\left(\EK\left(z\right)-i\EK\left(1-z\right)\right)\,,\quad \text{with}\quad z\to z+i0^{+}\,,
\end{align}
which establish linear relations between elliptic integrals with different reality domains. Therefore, by extending the set of building-blocks of the homogenous solutions to the elliptic integrals
\begin{align}
\EK\left(\omega_{\pm}\right)\,,\quad \EK\left(1-\omega_{\pm}\right)\,,\quad \EK\left(\frac{1}{\omega_{\pm}}\right)\,,\quad \EK\left(\frac{1}{1-\omega_{\pm}}\right)\,,\quad \EK\left(1-\frac{1}{\omega_{\pm}}\right)\,,
\end{align}
one can easily obtain, for each region $(a,b)$, a matrix of homogeneous solutions $G^{(a,b)}(x)$ with real entries. In the following we list, for each region, one possible choice of real solutions for the first master integral, which correspond to the first row of $G^{(a,b)}(x)$. As we have already observed, the other two rows can be obtained by applying the differential operators (\ref{eq:dx},~\ref{eq:d2x}) to the first one.
\begin{itemize}
\item{$-\infty<x<0$:}
\begin{align}
H_{1}^{(-\infty,0)}(x)=&\EK\left(\omega_{+}\right) \EK\left(\omega_{-}\right)\,,\nn
J_{1}^{(-\infty,0)}(x)=&\frac{1}{2}\left[\EK\left(\omega_{+}\right) \EK\left(1-\omega_{-}\right)+\EK\left(1-\omega_{+}\right) \EK\left(\omega_{-}\right)\right]\,,\nn
I_{1}^{(-\infty,0)}(x)=&\frac{1}{\sqrt{1-\omega_{-}}\sqrt{1-\omega_{+}}}\EK\left(\frac{1}{1-\omega_{+}}\right) \EK\left(\frac{1}{1-\omega_{-}}\right).
\label{eq:homI}
\end{align}
with Wronskian
\begin{align}
W^{(-\infty,0)}(x)=\frac{\pi ^3 x^3}{64 \sqrt{ (1-4x)^{3}(1-x)}}\,.
\end{align}
\item{$0<x<1/4$:}
\begin{align}
H_{1}^{(0,1/4)}(x)=&\frac{1}{2}\left[\EK\left(\omega_{+}\right) \EK\left(1-\omega_{-}\right)+\EK\left(1-\omega_{+}\right) \EK\left(\omega_{-}\right)\right]\,,\nn
J_{1}^{(0,1/4)}(x)=&-\frac{1}{2}\left[\EK\left(\omega_{+}\right) \EK\left(\omega_{-}\right)+\EK\left(1-\omega_{+}\right) \EK\left(1-\omega_{-}\right)\right]\,,\nn
I_{1}^{(0,1/4)}(x)=&\EK\left(\omega_{-}\right)\left[ \EK\left(\omega_{+}\right)+\frac{1}{\sqrt{\omega_{+}}}\EK\left(\frac{1}{\omega_{+}}\right) \right]\,,
\label{eq:homII}
\end{align}
with Wronskian
\begin{align}
W^{(0,1/4)}(x)=\frac{\pi ^3 x^3}{64  \sqrt{(1-4x)^{3}(1-x)}}\,.
\end{align}
\item{$1/4<x<1$:}
\begin{align}
H_{1}^{(1/4,1)}(x)=&\frac{1}{2}\left[\EK\left(\omega_{+}\right) \EK\left(\omega_{-}\right)+\EK\left(1-\omega_{+}\right) \EK\left(1-\omega_{-}\right)\right]\,,\nn
J_{1}^{(1/4,1)}(x)=&\frac{1}{\sqrt{\omega_{+}}}\EK\left(\omega_{-}\right) \EK\left(1-\frac{1}{\omega_{+}}\right)\,,\nn
I_{1}^{(1/4,1)}(x)=&-\frac{1}{\sqrt{1-\omega_{+}}\sqrt{\omega_{-}}}\EK\left(\frac{1}{1-\omega_{+}}\right)\EK\left(\frac{1}{\omega_{-}}\right)\,,
\label{eq:homIII}
\end{align}
with Wronskian
\begin{align}
W^{(1/4,1)}(x)=\frac{\pi ^3 x^3}{64 \sqrt{ (4 x-1)^{3}(1-x)}}\,.
\end{align}
\item{$1<x<\infty$:}
\begin{align}
H_{1}^{(1,\infty)}(x)=&\EK\left(\omega_{+}\right) \EK\left(\omega_{-}\right)\,,\nn
J_{1}^{(1,\infty)}(x)=&\EK\left(\omega_{+}\right) \EK\left(1-\omega_{-}\right)\,,\nn
I_{1}^{(1,\infty)}(x)=&-\frac{1}{3}\EK\left(1-\omega_{+}\right)\EK\left(1-\omega_{-}\right)\,,
\label{eq:homIV}
\end{align}
with Wronskian
\begin{align}
W^{(1,\infty)}(x)=\frac{\pi ^3 x^3}{64 \sqrt{(4 x-1)^{3}( x-1)}}\,.
\end{align}
\end{itemize}

\subsection{Analytic continuation of the homogenous solution}
Once real homogenous solutions $G^{(a,b)}(x)$ have been found in each region $(a,b)$, we must match them across the four singular points $x=0,1/4,1$ and $x=\pm \infty$ in order to analytically continue the homogenous solutions the whole range $-\infty<x<\infty$. Given the matrices $G^{(a,b)}(x)$  and $G^{(b,c)}(x)$ of real solutions defined in the adjacent intervals $(a,b)$ and $(b,c$), the analytic continuation amounts
 to define a matching matrix  $M^{(b)}$ ,
\begin{align}
G^{(b,c)}(x)=G^{(a,b)}(x)M^{(b)}\,,
\label{eq:defMatch}
\end{align}
which allows to continue $G^{(a,b)}(x)$ to the region $b<x<c$. The matrix $M^{(b)}$ can be obtained by assigning a 
small imaginary part to $x\to x-i0^{+}$ (the sign of which is inherited from the Feynman prescription $s\to s+i0^{+}$) 
and by equating the $x\to b^{+}$ limit of the two sides of \eqref{eq:defMatch}. \\
This procedure leads the four matching matrices
\begin{align}
M^{(0)}=&\left(
\begin{array}{ccc}
 0 & 1& -1 \\
 2 & -3i & 3i \\
 -i & -1/2 & 0 \\
\end{array}
\right)\,,\quad
M^{(1/4)}=\left(
\begin{array}{ccc}
 0 & 1& -1 \\
 -1 & -2i & 0 \\
 0 & -i & 0 \\
\end{array}
\right)\,,\nn
M^{(1)}=&\left(
\begin{array}{ccc}
 1 & 0 & -1/3 \\
 2 i & 3 & 2/3 i \\
 i/2& 0 & i/6 \\
\end{array}
\right)\,,\qquad
M^{(\infty)}=\left(
\begin{array}{ccc}
 1 & -i & -3 \\
 0 & -1/3 & 2i \\
 0 & 0 & -3 \\
\end{array}
\right)\,,
\label{eq:matchM}
\end{align}
which, consistently with Eq.~\eqref{eq:defMatch}, satisfy
\begin{align}
M^{(0)}M^{(1/4)}M^{(1)}M^{(\infty)}=1\,.
\end{align}
The limits of the homogenous solutions (\ref{eq:homI}~-\ref{eq:homIV}) close to the singular points, which have been used to obtain \eqref{eq:matchM}, can be easily calculated with the help of computer algebra system such as {\tt Mathematica} and, therefore, we will not write them down explicitly.  As an example, we will just list below the leading behaviour of the homogenous solutions \eqref{eq:homIV} at the end-points of the region $(1,\infty)$, which have been also used in Section~\ref{sec:inhomsol} in order to fix the boundary constants of the inhomogeneous solution.\\

The limit of $G^{(1,\infty)}(x)$ for $x\to 1^{+}$ is
\begin{align}
\lim_{x\to 1^{+}}H^{(1,\infty)}_{1}(x)=&\EK\left(r_{+}\right) \EK\left(r_{-}\right)+\mathcal{O}\left(\sqrt{x-1}\right)\,,\nn
\lim_{x\to 1^{+}}J^{(1,\infty)}_{1}(x)=&\EK\left(r_{+}\right) \EK\left(r_{+}\right)+\mathcal{O}\left(\sqrt{x-1}\right)\,,\nn
\lim_{x\to 1^{+}}I^{(1,\infty)}_{1}(x)=&-\frac{1}{3} \EK\left(r_{+}\right)\EK\left(r_{-}\right)+\mathcal{O}\left(\sqrt{x-1}\right)\,,\nn
\lim_{x\to 1^{+}}H^{(1,\infty)}_{2}(x)=&\frac{1}{26 \sqrt{x-1}}\bigg(\EE\left(r_{-}\right) \left(6 \EE\left(r_{+}\right)+\left(\sqrt{3}-9\right) \EK\left(r_{+}\right)\right)\nn
&-\EK\left(r_{-}\right) \left(\left(9+\sqrt{3}\right) \EE\left(r_{+}\right)-6
   \EK\left(r_{+}\right)\right)\bigg)+\mathcal{O}\left(\sqrt{x-1}\right)\,,\nn
\lim_{x\to 1^{+}}J^{(1,\infty)}_{2}(x)=&\frac{1}{6}\left(\left(\sqrt{3}-3\right) \EK\left(r_{+} \right)^2-2 \left(\sqrt{3}-3\right)
   \EK\left(r_{+} \right) \EE\left(r_{+} \right)-6 \EE\left(r_{+}\right)^2\right)+\mathcal{O}\left(\sqrt{x-1}\right)\,\nn
\lim_{x\to 1^{+}}I^{(1,\infty)}_{2}(x)=&\frac{1}{12 \sqrt{x-1}}	\left(\EK\left(r_{+} \right) \left(\EK\left(r_{-}\right)-\EE\left(r_{-} \right)\right)-\EK\left(r_{-} \right) \EE\left(r_{+}\right)\right)+\nn
   &\frac{1}{18} \left(\left(3+\sqrt{3}\right) \EK\left(r_{-} \right) \EE\left(r_{+}\right)-\EE\left(r_{-} \right) \left(\left(\sqrt{3}-3\right) \EK\left(r_{+}\right)+6 \EE\left(r_{+} \right)\right)\right)+\mathcal{O}\left(\sqrt{x-1}\right)\,,\nn
   \lim_{x\to 1^{+}}H^{(1,\infty)}_{3}(x)=&\frac{1}{8 \sqrt{x-1}}\left(\EK\left(r_{-} \right) E\left(r_{+} \right)+\EK\left(r_{+}\right) \left(\EE\left(r_{-} \right)-\EK\left(r_{-}\right)\right)\right)+\nn
   &{1}{36} \left(\EK\left(r_{-} \right) \left(4 \left(3+\sqrt{3}\right) E\left(r_{+}\right)-3 \EK\left(r_{+} \right)\right)-4 \EE\left(r_{-}\right) \left(\left(\sqrt{3}-3\right) \EK\left(r_{+} \right)+6 \EE\left(r_{+}\right)\right)\right)\nn
   &+\mathcal{O}\left(\sqrt{x-1}\right)\,,\nn
\lim_{x\to 1^{+}}J^{(1,\infty)}_{3}(x)=&\frac{1}{36} \left(\left(9-4 \sqrt{3}\right) \EK\left(r_{+} \right)^2+8 \left(\sqrt{3}-3\right)
   \EK\left(r_{+} \right) \EK\left(r_{+} \right)+24 \EE\left(r_{+}\right)^2\right)+\mathcal{O}\left(\sqrt{x-1}\right)\,,\nn
\lim_{x\to 1^{+}}I^{(1,\infty)}_{3}(x)=&\frac{1}{24 \sqrt{x-1}}\left(\EK\left(r_{-} \right) E\left(r_{+} \right)+\EK\left(r_{+}\right) \left(\EE\left(r_{-} \right)-\EK\left(r_{-}\right)\right)\right)+\nn
   &\frac{1}{108} \big(4 \EE\left(r_{-}\right) \left(6 \EE\left(r_{+}\right)+\left(\sqrt{3}-3\right) \EK\left(r_{+}\right)\right)\nn
   &+\EK\left(r_{-}\right) \left(3 \EK\left(r_{+}\right)-4 \left(3+\sqrt{3}\right) \EE\left(r_{+}\right)\right)\big)+\mathcal{O}\left(\sqrt{x-1}\right)\,.\nn
   \label{eq:exp1p}
 \end{align}
where we have defined 
\begin{align}
r_{\pm}\equiv\lim_{x\to 1^{+}}\omega_{\pm}=\frac{2\pm \sqrt{3}}{4}.
\end{align}
The leading behaviour of $G^{(1,\infty)}(x)$ for $x\to +\infty$ is instead
\begin{align}
\lim_{x\to +\infty}H^{(1,\infty)}_{1}(x)=&\frac{\pi^2}{4}+\mathcal{O}\left(1/x\right)\,,\nn
\lim_{x\to +\infty}J^{(1,\infty)}_{1}(x)=&\frac{3}{4}\pi\left(4\ln{2}-\ln\left(1/x\right)\right)+\mathcal{O}\left(1/x\right)\,,\nn
\lim_{x\to +\infty}I^{(1,\infty)}_{1}(x)=&\frac{1}{2}\left(\ln\left(1/x\right)-4\ln{2}\right)^2+\mathcal{O}\left(1/x\right)\,,\nn
\lim_{x\to +\infty}H^{(1,\infty)}_{2}(x)=&-\frac{\pi^2}{16}+\mathcal{O}\left(1/x\right)\,,\nn
\lim_{x\to +\infty}J^{(1,\infty)}_{2}(x)=&\frac{3\pi}{16}\left(1-4\ln{2}+\ln\left(1/x\right)\right)+\mathcal{O}\left(1/x\right)\,,\nn
\lim_{x\to +\infty}I^{(1,\infty)}_{2}(x)=&
 \frac{1}{16} (4 \ln{2}-\ln\left(1/x\right)-2) (4 \ln\left(1/x\right)-\ln\left(1/x\right)) +\mathcal{O}\left(1/x\right)\,,\nn
\lim_{x\to +\infty}H^{(1,\infty)}_{3}(x)=&-\frac{\pi^2}{64}+\mathcal{O}\left(1/x\right)\,,\nn
\lim_{x\to +\infty}J^{(1,\infty)}_{3}(x)=&\frac{3\pi}{128}   \left(-2 \ln \left(1/x\right)-3+8 \ln{2}\right)+\mathcal{O}\left(1/x\right)\,,\nn
\lim_{x\to +\infty}I^{(1,\infty)}_{3}(x)=&\frac{x}{24}
-\frac{1}{64} (4 \ln{2}-\ln\left(1/x\right)-3) (4 \ln{2}-\ln{2})\ln\left(1/x\right)+\mathcal{O}\left(1/x\right)\,.
   \label{eq:expinf}
   \end{align}  

\subsection{Analytic continuation of the inhomogeneous solution}
We are finally in the position to analytically continue the inhomogeneous solutions to arbitrary values of $x$.
As it is explicitly shown by Eq.~\eqref{eq:Isol}, the inhomogeneous solution is defined, for $x>1$, in terms of integrals of the functions $\mathcal{R}^{(1,\infty)}_{i}(x)$ which, in turn, depend on the homogenous solutions $G^{(1,\infty)}(x)$ and on their Wronskian $W^{(1,\infty)}(x)$. Therefore, in order to extend the integral representation \eqref{eq:Isol} to the other kinematic regions, it is sufficient to analytically continue the elements of $G^{(1,\infty)}(x)$ appearing in the definition \eqref{rotatedM1} of  $\mathcal{R}^{(1,\infty)}_{i}(x)$, by making use of the matching matrices $M^{(b)}$, as prescribed by Eq.~\eqref{eq:defMatch}. In this way all imaginary parts (whenever they are present) are made explicit and, as a result, we obtain a representation of the solution which involves, for any $x$, the evaluation real integrals only.
\\

We start by considering the analytic continuation to $1/4<x<1$. 
The matching matrix $M^{(1)}$, which has been defined in Eq.~\eqref{eq:matchM}, 
can be used in order to express the homogeneous solutions $G^{(1,\infty)}$ 
in terms of the set of real solutions defined in $(1/4,1)$,
\begin{align}
G^{(1,\infty)}(x)=G^{(1/4,1)}(x)M^{(1)}.
\label{eq:GMatch1}
\end{align}
In addition, it is easy to see that, with the Feynman prescription $x\to x-i\eps$, the Wronskian is analytically continued for $1/4<x<1$ as 
\begin{align}
W^{(1,\infty)}=\frac{\pi ^3 x^3}{64  \sqrt{(4 x-1)^{3}(x-1)}}=\frac{i\pi ^3 x^3}{64 \sqrt{ (1-4x)^{3}(1-x)}}=iW^{(1/4,1)}\,.
\label{eq:Wcont1}
\end{align}
By acting with Eq.s~\eqref{eq:GMatch1} and \eqref{eq:Wcont1} on the inhomogeneous solution \eqref{eq:matchM}, we can write the master integrals in region $(1/4,1)$ in terms of individually real-valued integrals as
\begin{align}
\I_{1}^{(0)}(x)=&H^{(1/4,1)}_{1}(x) \left(c_{1}^{(0)}+\int_{x}^1  \frac{dt}{1-4t}\mathcal{R}^{(1/4,1)}_{1}(t)\right)+
J^{(1/4,1)}_{1}(x)\left( c_{2}^{(0)}+\int_{x}^1 \frac{dt}{1-4t}\mathcal{R}^{(1/4,1)}_{2}(t)\right)\nn
&+I^{(1/4,1)}_{1}(x) \left(c_{3}^{(0)}+\int_{x}^1\frac{dt}{1-4t}\mathcal{R}^{(1/4,1)}_{3}(t)\right)
 \,,\nn
\I_{2}^{(0)}(x)=&H^{(1/4,1)}_{2}(x) \left(c_{1}^{(0)}+\int_{x}^1 \frac{dt}{1-4t}\mathcal{R}^{(1/4,1)}_{1}(t)\right)+J^{(1/4,1)}_{2}(x)\left( c_{2}^{(0)}+\int_{x}^1 \frac{dt}{1-4t}\mathcal{R}^{(1/4,1)}_{2}(t)\right)\nn
&+I^{(1/4,1)}_{2}(x) \left(c_{3}^{(0)}+\int_{x}^1 \frac{dt}{1-4t}\mathcal{R}^{(1/4,1)}_{3}(t)\right)\,,\nn
\I_{3}^{(0)}(x)=&H^{(1/4,1)}_{3}(x) \left(c_{1}^{(0)}+\int_{x}^1 \frac{dt}{1-4t}\mathcal{R}^{(1/4,1)}_{1}(t)\right)+J^{(1/4,1)}_{3}(x)\left( c_{2}^{(0)}+\int_{x}^1 \frac{dt}{1-4t}\mathcal{R}^{(1/4,1)}_{2}(t)\right)\nn
&+I^{(1/4,1)}_{3}(x)\left(c_{3}^{(0)}+\int_{x}^1 \frac{dt}{1-4t}\mathcal{R}^{(1/4,1)}_{3}(t)\right)\,,
\label{eq:IsolrII}
\end{align}
where $\mathcal{R}^{(1/4,1)}_{i}(x)$ are combinations of homogenous solutions defined by
\begin{align}
\mathcal{R}^{(1/4,1)}_{1}(x) =& \frac{i}{4W^{(1/4,1)}(x)}\bigg[2H_{1}^{(1/4,1)}(x)J_{2}^{(1/4,1)}(x)-2H_{2}^{(1/4,1)}(x)J_{1}^{(1/4,1)}(x)+\nn
&i \left(I_{2}^{(1/4,1)}(x)J_{1}^{(1/4,1)}(x)-I_{1}^{(1/4,1)}(x)J_{2}^{(1/4,1)}(x)\right)\bigg]\,,\nn
\mathcal{R}^{(1/4,1)}_{2}(x) =& -\frac{1}{6W^{(1/4,1)}(x)}\bigg[H_{2}^{(1/4,1)}(x)\left(I_{1}^{(1/4,1)}(x)+4J_{1}^{(1/4,1)}(x)\right)-\nn
&H_{1}^{(1/4,1)}(x)\left(I_{2}^{(1/4,1)}(x)+4J_{2}^{(1/4,1)}(x)\right)\bigg]\,,\nn
\mathcal{R}^{(1/4,1)}_{3}(x) =& \frac{3i}{4W^{(1/4,1)}(x)}\bigg[2H_{1}^{(1/4,1)}(x)J_{2}^{(1/4,1)}(x)-2H_{2}^{(1/4,1)}(x)J_{1}^{(1/4,1)}(x)+\nn
&i \left(I_{1}^{(1/4,1)}(x)J_{2}^{(1/4,1)}(x)-I_{2}^{(1/4,1)}(x)J_{1}^{(1/4,1)}(x)\right)\bigg]\,.\nn
\end{align}

We can now continue the solution to $0<x<1/4$, where the master integrals develop an imaginary part. The region $(0,1/4)$ must be linked to $(1,\infty)$ by passing through  region $(1/4,1)$. This means that, according to the definition \eqref{eq:defMatch}, the homogenous solutions $G^{(1,\infty)}(x)$ are continued in terms of the real-valued solutions defined for  $0<x<1/4$ as
\begin{align}
G^{(1,\infty)}(x)=G^{(0,1/4)}(x)M^{(1/4)}M^{(1)}\,,
\label{eq:GMatch1o4}
\end{align}
where $M^{(1/4)}$ and $M^{(1)}$ are the matching matrices given in Eq.~\eqref{eq:matchM}. In this case, Wronskian is trivially continued, 
\begin{align}
W^{(1,\infty)}=\frac{\pi ^3 x^3}{64 \sqrt{ (4 x-1)^{3}(x-1)}}=\frac{\pi ^3 x^3}{64 \sqrt{(1-4 x)^{3}(1-x)}}=W^{(0,1/4)}\,,
\label{eq:Wcont1o4}
\end{align}
and by making use of Eq.s~\eqref{eq:GMatch1o4} and \eqref{eq:Wcont1o4} we can write master integrals in region $(0,1/4)$ as
\begin{align}
\I_{1}^{(0)}(x)=&H^{(0,1/4)}_{1}(x)\left(b_{1}^{(0)}+\int_{1/4}^x \frac{dt }{1-4t}\mathcal{R}^{(0,1/4)}_{1}(t)\right)+
J^{(0,1/4)}_{1}(x)\left( b_{2}^{(0)}+\int_{1/4}^x \frac{dt }{1-4t}\mathcal{R}^{(0,1/4)}_{2}(t)\right)+\nn
&I^{(0,1/4)}_{1}(x) \left(b_{3}^{(0)}+\int_{1/4}^x \frac{dt }{1-4t}\mathcal{R}^{(0,1/4)}_{3}(t)\right)
 \,,\nn
\I_{2}^{(0)}(x)=&H^{(0,1/4)}_{2}(x) \left(b_{1}^{(0)}+\int_{1/4}^x \frac{dt }{1-4t}\mathcal{R}^{(0,1/4)}_{1}(t)\right)+J^{(0,1/4)}_{2}(x)\left(b_{2}^{(0)}+\int_{1/4}^x \frac{dt}{1-4t}\mathcal{R}^{(0,1/4)}_{2}(t)\right)\nn
&+I^{(0,1/4)}_{2}(x) \left(b_{3}^{(0)}+\int_{1/4}^x \frac{dt}{1-4t}\mathcal{R}^{(0,1/4)}_{3}(t)\right)\,,\nn
\I_{3}^{(0)}(x)=&H^{(0,1/4)}_{3}(x) \left(b_{1}^{(0)}+\int_{1/4}^x \frac{dt}{1-4t}\mathcal{R}^{(0,1/4)}_{1}(t)\right)+J^{(0,1/4)}_{3}(x)\left(b_{2}^{(0)}+\int_{1/4}^xd\frac{dt }{1-4t}\mathcal{R}^{(0,1/4)}_{2}(t)\right)\nn
&+I^{(0,1/4)}_{3}(x)\left( b_{3}^{(0)}+\int_{1/4}^x \frac{dt }{1-4t}\mathcal{R}^{(0,1/4)}_{3}(t)\right) \,,
\label{eq:IsolrII}
\end{align}
where the integration constants $b_{i}^{(0)}$ are defined by
\begin{align}
b_{i}^{(0)}=c_{i}^{(0)}+\int_{1/4}^1  \frac{dt }{1-4t}\mathcal{R}^{(1/4,1)}_{i}(t)\,,
\end{align}
and $\mathcal{R}^{(0,1/4)}_{i}(x)$ are the combinations of homogenous solutions
\begin{align}
\mathcal{R}^{(0,1/4)}_{1}(x) =& \frac{1}{4W^{(0,1/4)}(x)}\bigg[H_{1}^{(0,1/4)}(x)\left(I_{2}^{(0,1/4)}(x)+4J_{2}^{(0,1/4)}(x)\right)-\nn
&H_{2}^{(0,1/4)}(x)\left(I_{1}^{(0,1/4)}(x)+4J_{1}^{(0,1/4)}(x)\right)\nn
&-2i\left(I_{2}^{(0,1/4)}(x)J_{1}^{(0,1/4)}(x)-I_{1}^{(0,1/4)}(x)J_{2}^{(0,1/4)}(x)\right)\bigg]\,,\nn
\mathcal{R}^{(0,1/4)}_{2}(x) =& \frac{1}{6W^{(0,1/4)}(x)}\bigg[4I_{2}^{(0,1/4)}(x)J_{1}^{(0,1/4)}(x)-4I_{1}^{(0,1/4)}(x)J_{2}^{(0,1/4)}(x)\nn
&+3i\left(H_{2}^{(0,1/4)}(x)J_{1}^{(0,1/4)}(x)-H_{1}^{(0,1/4)}(x)J_{2}^{(0,1/4)}(x)\right)\bigg]\,,\nn
\mathcal{R}^{(0,1/4)}_{3}(x) =& \frac{3}{4W^{(0,1/4)}(x)}\bigg[I_{1}^{(0,1/4)}(x)H_{2}^{(0,1/4)}(x)-I_{2}^{(0,1/4)}(x)H_{1}^{(0,1/4)}(x)\nn
&+2i\left(I_{2}^{(0,1/4)}(x)J_{1}^{(0,1/4)}(x)-I_{1}^{(0,1/4)}(x)J_{2}^{(0,1/4)}(x)\right)\bigg]\,.
\end{align}
Finally, the expression of the master integrals in the Euclidean region $x<0$ can be obtained by matching the homogenous solutions at infinity, according to Eq.~\eqref{eq:defMatch},
\begin{align}
G^{(1,\infty)}(x)=G^{(-\infty,0)}(x)\left(M^{(\infty)}\right)^{-1},
\label{eq:GMatchoo}
\end{align}
with the matching matrix $M^{(\infty)}$ defined in Eq.\eqref{eq:matchM}. The Wronskian can be directly continued to negative values of $x$,
\begin{align}
W^{(1,\infty)}=\frac{\pi ^3 x^3}{64 \sqrt{ (4 x-1)^{3}(x-1)}}=\frac{\pi ^3 x^3}{64  \sqrt{(1-4 x)^{3}(1-x)}}=W^{(-\infty,0)}\,,
\label{eq:Wcontoo}
\end{align}
and by acting again with \eqref{eq:GMatchoo} and \eqref{eq:Wcontoo} on Eq.~\eqref{eq:Isol}, we obtain the expression of the master integrals in region $(-\infty,0)$,
\begin{align}
\I_{1}^{(0)}(x)&=H^{(-\infty0)}_{1}(x)\left(d_{1}^{(0)}+\int_0^{-1/x} \frac{dy}{y(1+4y)}\mathcal{R}^{(-\infty,0)}_{1}\left(1/y\right)\right)+\nn
&J^{(-\infty,0)}_{1}(x)\int_0^{-1/x}\frac{dy}{y(1+4y)}\mathcal{R}^{(-\infty,0)}_{2}\left(1/y\right)+I^{(-\infty,0)}_{1}(x) \int_0^{-1/x}\frac{dy}{y(1+4y)}\mathcal{R}^{(-\infty,0)}_{3}\left(1/y\right)
 \,,\nn
\I_{2}^{(0)}(x)&=H^{(-\infty,0)}_{2}(x) \left(d_{1}^{(0)}+\int_0^{-1/x}\frac{dy}{y(1+4y)}\mathcal{R}^{(-\infty,0)}_{1}\left(1/y\right)\right)+\nn
&J^{(-\infty,0)}_{2}(x)\int_0^{-1/x}\frac{dy}{y(1+4y)}\mathcal{R}^{(-\infty,0)}_{2}\left(1/y\right)+I^{(-\infty,0)}_{2}(x) \int_0^{-1/x}\frac{dy}{y(1+4y)}\mathcal{R}^{(-\infty,0)}_{3}\left(1/y\right)\,,\nn
\I_{3}^{(0)}(x)&=H^{(-\infty,0)}_{3}(x) \left(d_{1}^{(0)}+\int_0^{-1/x}\frac{dy}{y(1+4y)}\mathcal{R}^{(-\infty,0)}_{1}(t)\right)+\nn
&J^{(-\infty,0)}_{3}(x)\int_0^{-1/x}\frac{dy}{y(1+4y)}\mathcal{R}^{(-\infty,0)}_{2}\left(1/y\right)+I^{(-\infty,0)}_{3}(x) \int_0^{-1/x} \frac{dy}{y(1+4y)}\mathcal{R}^{(-\infty,0)}_{3}\left(1/y\right) \,,
\label{eq:IsolrI}
\end{align}
where $d_{1}^{(0)}$ is the integration constant
\begin{align}
d_{1}^{(0)}=c_{1}^{(0)}+\int_{0}^1 \frac{dy}{y(y-4)}\mathcal{R}^{(1,\infty)}_{1}\left(1/y\right)\,,
\end{align}
and $\mathcal{R}^{(-\infty,0)}_{i}(x)$ are the combinations of homogenous solutions
\begin{align}
\mathcal{R}^{(-\infty,0)}_{1}(x) =& \frac{1}{2W^{(-\infty,0)}(x)}\bigg[J_{1}^{(-\infty,0)}(x)\left(3H_{2}^{(-\infty,0)}(x)+I_{2}^{(-\infty,0)}(x)\right)\nn
&-J_{2}^{(-\infty,0)}(x)\left(3H_{1}^{(-\infty,0)}(x)+I_{1}^{(-\infty,0)}(x)\right)\nn
&+i\left(H_{1}^{(-\infty,0)}(x)I_{2}^{(-\infty,0)}(x)-H_{2}^{(-\infty,0)}(x)I_{1}^{(-\infty,0)}(x)\right)\bigg]\,,\nn
\mathcal{R}^{(-\infty,0)}_{2}(x) =&\frac{1}{6W^{(-\infty,0)}(x)}\bigg[\left(H_{1}^{(-\infty,0)}(x)I_{2}^{(-\infty,0)}(x)-H_{2}^{(-\infty,0)}(x)I_{1}^{(-\infty,0)}(x)\right)\nn
&+6i\left(H_{1}^{(-\infty,0)}(x)J_{2}^{(-\infty,0)}(x)-H_{2}^{(-\infty,0)}(x)J_{1}^{(-\infty,0)}(x)\right)\bigg]\,,\nn
\mathcal{R}^{(1/4,1)}_{3}(x) =& \frac{1}{6W^{(-\infty,0)}(x)}\bigg[H_{2}^{(-\infty,0)}(x)J_{1}^{(-\infty,0)}(x)-H_{1}^{(-\infty,0)}(x)J_{2}^{(-\infty,0)}(x)\bigg]\,.
\end{align}

We stress here that similar results in the region $(-\infty,0)$ could be obtained by matching the solutions in $x=0$.
This point, nevertheless, is more delicate, due to the divergence in the Wronskian, i.e. $1/W(x) \sim 1/x^3$ as $x \to 0$, and we
preferred for this reason to continue passing through $x=\pm \infty$.

\section{Proof of Eq.s~\eqref{eq:rescut1}-\eqref{eq:rescut2}}
\label{ap:1} \setcounter{equation}{0} 
\numberwithin{equation}{section} 
In this appendix we give a brief derivation of Eq.s~\eqref{eq:rescut1}-\eqref{eq:rescut2}.
We first reproduce the proof of Eq.~\eqref{eq:rescut2}, which was first presented in~\cite{Bailey:2008ib} (see Eq.(33) therein), and then we use similar arguments to derive Eq.~\eqref{eq:rescut1}.\\
The evaluation of Eq.~\eqref{eq:fincut2}  requires the study of the integral 
\begin{align}
\pi \int_0^\infty \frac{dt}{\sqrt{( t^2+(a+b)^2)( t^2 +c^2)  }} 
\EK\left( \frac{t^2+(a-b)^2}{t^2+(a+b)^2} \right)
\label{eq:id1}
\end{align}
The analytic expression of \eqref{eq:id1} can be obtained by first studying the following auxiliary integral
\begin{align}
I_1(\omega)=\frac{2}{\pi}\int_{0}^{\infty}dtdz_1dz_2\,K_0(az_1)K_0(bz_1)K_0\left(cz_2\right)\cos{(tz_1)}\cos{((\omega+t)z_2)}\,,
\label{eq:aux}
\end{align}
where $K_0(x)$ is the modified Bessel function of the second kind,
\begin{align}
K_0(x)=\int_{0}^{\infty}dt\frac{\cos{(xt)}}{\sqrt{t^2+1}}\,,
\end{align}
which satisfies the identity
\begin{align}
\label{eq:root}
\int_{0}^{\infty}dt K_0(at)\cos({\omega t})=\frac{2}{\pi}\frac{1}{\sqrt{a^2+\omega^2}}\,.
\end{align} 
Eq.~\eqref{eq:root} allows to  trivially perform the integration over $z_2$,
\begin{align}
I_1(\omega)=\int_{0}^{\infty}dt\frac{1}{\sqrt{(\omega+t)^2+c^2}}
\int_{0}^{\infty}dz_1\,K_0(az_1)K_0(bz_1)\cos{(tz_1)}\,.
\end{align}
The integral over $z_1$ is now in standard form (see for instance Eq.2.16.36.2 of \cite{Prudnikov}) 
and it can be evaluated in terms of an elliptic integral of the first kind,
\begin{align}
I_1(\omega)=\pi \int_{0}^{\infty}\frac{dt}{\sqrt{t^2+(a+b)^2}\sqrt{(\omega+t)^2+c^2}}\EK\left(\frac{t^2+(a-b)^2}{t^2+(a+b)^2}\right)\,,
\end{align}
from which we immediately see that Eq.~\eqref{eq:root} corresponds to the value of the auxiliary integral~\eqref{eq:aux} 
at  $\omega=0$. In order to evaluate $I_{1}(0)$, we go back to Eq.~\eqref{eq:aux} we start by performing the $dt$ integration
, for which we can use the 
distribution identity
\begin{align}
\int_{0}^{\infty} dt \cos{(tz_1)}\cos{((\omega+t)z_2)}=\frac{\pi}{2}\cos({\omega z_1})\ (\delta(z_1-z_2)+\cos({\omega z_1})\delta(z_1+z_2))\,.
\label{eq:delta} 
\end{align}
The term proportional to $\delta(z_1+z_2)$ in the right-hand-side of Eq.~\eqref{eq:delta} 
has no support in the region where $I_{1}(\omega)$ is defined, therefore we have
\begin{align}
I_1(\omega)=\int_{0}^{\infty}dz_1\,K_0(az_1)K_0(bz_1)K_0\left(cz_2\right)\cos{(\omega z_1)}\,,
\end{align}
which, if we set  $\omega=0$, reduces to
\begin{align}
I_1(0)=\int_{0}^{\infty}dz_1\,K_0(az_1)K_0(bz_1)K_0\left(cz_2\right)\,.
\end{align}
This last integral is connected to the master formula (see Eq(3.3) of \cite{PLMS:PLMS0037})
\begin{align}
\int_{0}^{\infty}dt I_\mu(at)K_0(bt)K_0(ct)dt=\frac{1}{4c}W_{\mu}(k_{+})W_{\mu}(k_{-}),
\label{eq:bailey}
\end{align}
where $I_{\mu}(z)$ is the modified Bessel function of the first kind, the function $W_{\mu}(k)$ is related to associated Legendre polynomial $P_{-1/2}^{\mu/2}$,
\begin{align}
W_{\mu}(k)=\frac{\sqrt{\pi}\Gamma\left(\frac{1+\mu}{2}\right)}{(1-k^2)^{1/4}}P_{-1/2}^{\mu/2}\left(\frac{2k^2}{2\sqrt{1-k^2}}\right)\,,
\end{align}
and the arguments $k_{\pm}$ are defined by
\begin{align}
k_{\pm}=\frac{\sqrt{(c+a)^2-b^2}\pm \sqrt{(c-a)^2-b^2}}{2c}.
\end{align}
The expansion of Eq.\eqref{eq:bailey} around $\mu=0$ allow us to express a set of integrals of three Bessel functions as a product of two complete elliptic integrals. In fact, by making use of
\begin{align}
\nonumber
I_{\mu}(x)=&I_{0}(x)-\mu K_{0}(x)+\mathcal{O}(\mu^2),\\
W_{\mu}(k)=&2\EK(k^2)-\mu \pi \EK(1-k^2)+\mathcal{O}(\mu^2),
\end{align}
one can easily check that that Eq.~\eqref{eq:bailey} implies
\begin{align}
\label{eq:IKK}
\int_{0}^{\infty}dt I_0(at)K_0(bt)K_0(ct)dt=&\frac{1}{c}\EK(k_{-}^2)\EK(k_{+}^2) ,\\
\label{eq:KKK}
\int_{0}^{\infty}dt K_0(at)K_0(bt)K_0(ct)dt=&\frac{\pi}{2c}\left(\EK(k_{-}^2)\EK(1-k_{+}^2)+\EK(k_{+}^2)\EK(1-k_{-}^2)\right)\,.
\end{align}
Thanks to Eq.~\eqref{eq:KKK} we can finally evaluate $I_{1}(0)$,
\begin{align}
I_1(0)=\frac{\pi}{2c}\left(\EK(k_{-}^2)\EK(1-k_{+}^2)+\EK(k_{+}^2)\EK(1-k_{-}^2)\right)\,,
\label{eq:11}
\end{align}
which proves Eq.~\eqref{eq:rescut2}.\\

The proof of Eq.~\eqref{eq:rescut2}, which requires the evaluation of the integral
\begin{align}
 \int_0^\infty \frac{dt}{\sqrt{( t^2+(a+b)^2)( t^2 +c^2)  }} 
\EK\left( \frac{2ab}{t^2+(a+b)^2} \right)\,,
\label{eq:id2}
\end{align}
proceeds along the same lines. We start from the auxiliary integral
\begin{align}
I_2(\omega)=\frac{2}{\pi}\int_{0}^{\infty}dtdz_1dz_2\,I_0(az_1)K_0(bz_1)K_0\left(cz_2\right)\cos{(tz_1)}\cos{((\omega+t)z_2)}\,,
\label{eq:aux2}
\end{align}
which, by using in order \eqref{eq:root}, becomes
\begin{align}
\mathcal{I}_2(\omega)=\int_{0}^{\infty}dt\frac{1}{\sqrt{(\omega+t)^2+c^2}}
\int_{0}^{\infty}dz_1\,I_0(az_1)K_0(bz_1)\cos{(tz_1)}.
\end{align}
As in the previous case, the integral over $z_1$ can be evaluated in terms of an elliptic integral of the first kind (see for instance Eq.2.16.36.2 of \cite{Prudnikov}),
\begin{align}
I_2(\omega)= \int_{0}^{\infty}\frac{dt}{\sqrt{t^2+(a+b)^2}\sqrt{(\omega+t)^2+c^2}}\EK\left(\frac{t^2+(a-b)^2}{t^2+(a+b)^2}\right),
\end{align}
from which we see that Eq.~\eqref{eq:id2} corresponds to $I_2(0)$. Therefore, in order to determine the value of the auxiliary integral at zero, we first make use of \eqref{eq:delta} in Eq.~\eqref{eq:aux2} in order to integrate over $t$
\begin{align}
I_2(\omega)=\int_{0}^{\infty}dz_1\,I_0(az_1)K_0(bz_1)K_0\left(cz_2\right)\cos{(\omega z_1)}.
\end{align}
and then, after setting $\omega=0$,
\begin{align}
I_2(0)=\int_{0}^{\infty}dz_1\,I_0(az_1)K_0(bz_1)K_0\left(cz_2\right)\,,
\end{align}
 we make use Eq.~\eqref{eq:IKK} and obtain
\begin{align}
I_2(0)=\frac{1}{c}\EK(k_{-}^2)\EK(k_{+}^2)\,,
\label{eq:21}
\end{align}
which proves Eq.~\eqref{eq:rescut1}.

\bibliographystyle{bibliostyle}   
\bibliography{Biblio}
\end{document}